\documentclass[prc,twocolumn,twoside,showpacs,floatfix,superscriptaddress]{revtex4}

\usepackage{graphicx,color,rotating}
\usepackage{amsmath,amssymb}
\usepackage{dcolumn}
\usepackage{times,txfonts}
\usepackage{pifont}

\usepackage{multirow}

\usepackage{mathrsfs}

\usepackage{dsfont}

\usepackage{pstricks}
\usepackage{pst-text}
\usepackage{pst-coil}
\usepackage{pst-node}
\usepackage{pstricks-add}

\newcommand{\symboldiamond}[1][black]{{\color{#1}\hspace{-1pt}\footnotesize\begin{turn}{45} $\blacksquare$ \end{turn}}}
\newcommand{\symboltriangle}[1][black]{{\color{#1}$\blacktriangle$}}

\newcommand{\symbolcircle}[1][black]{{\color{#1}\large$\bullet$}}

\definecolor{FGViolet}{rgb}{0.61,0.32,0.61}
\definecolor{FGDarkBlue}{rgb}{0,0,0.6}
\definecolor{FGBlue}{rgb}{0,0,0.8}
\definecolor{FGLightBlue}{rgb}{0.2, 0.6, 0.8}
\definecolor{FGGreen}{rgb}{0.2,0.7,0.2}
\definecolor{FGLightGreen}{rgb}{0.4,1,0.4}
\definecolor{FGYellow}{rgb}{1,0.95,0}
\definecolor{FGOrange}{rgb}{0.95,0.5,0.1}
\definecolor{FGRed}{rgb}{0.8,0,0}
\definecolor{FGWhite}{rgb}{1,1,1}
\definecolor{FGLightGray}{rgb}{0.8,0.8,0.8}
\definecolor{FGGray}{rgb}{0.5,0.5,0.5}
\definecolor{FGDarkGray}{rgb}{0.3,0.3,0.3}
\definecolor{FGBlack}{rgb}{0,0,0}

\newcommand{\fm}{\ensuremath{ {\rm fm}^4 \ }}
\newcommand{\fmNoSpace}{\ensuremath{ {\rm fm}^4}}
\newcommand{\elem}[2]{\ensuremath{{}^{#2}\text{#1}}}
\newcommand{\emax}{\ensuremath{e_{\text{max}}}}

\newcommand{\ERef}{E_{\text{ref}}}
\newcommand{\mWME}[2]{w^{#1}_{#2}}

\newcommand{\beq}{\begin{equation}}
\newcommand{\eeq}{\end{equation}}
\newcommand{\half}{\mbox{$\frac{1}{2}$}}
\newcommand{\qua}{\mbox{$\frac{1}{4}$}}

\newcommand{\six}{\mbox{$\frac{1}{6}$}}

\newcommand{\tfour}{\mbox{$\frac{1}{24}$}}
\newcommand{\tsix}{\mbox{$\frac{1}{36}$}}
\newcommand{\onetwenty}{\mbox{$\frac{1}{120}$}}

\newcommand\helem[2]{\overline{h}{}^{#1}_{#2}({\rm 2B})}
\newcommand\welem[2]{\overline{w}{}^{#1}_{#2}}

\begin{document}


\title{Extension of coupled-cluster theory with a non-iterative treatment of connected triply excited clusters to three-body Hamiltonians}

\author{Sven Binder}
\thanks{Corresponding author}
\email{sven.binder@physik.tu-darmstadt.de}
\affiliation{Institut f\"ur Kernphysik, Technische Universit\"at Darmstadt, 64289 Darmstadt, Germany}

\author{Piotr Piecuch}
\email{piecuch@chemistry.msu.edu}
\affiliation{Department of Chemistry, Michigan State University, East Lansing, Michigan 48824, USA}

\author{Angelo Calci}
\email{angelo.calci@physik.tu-darmstadt.de}
\affiliation{Institut f\"ur Kernphysik, Technische Universit\"at Darmstadt, 64289 Darmstadt, Germany}

\author{Joachim Langhammer}
\email{joachim.langhammer@physik.tu-darmstadt.de}
\affiliation{Institut f\"ur Kernphysik, Technische Universit\"at Darmstadt, 64289 Darmstadt, Germany}

\author{Petr Navr\'atil}
\email{navratil@triumf.ca}
\affiliation{TRIUMF, 4004 Wesbrook Mall, Vancouver, British Columbia, V6T 2A3 Canada}

\author{Robert Roth}
\email{robert.roth@physik.tu-darmstadt.de}
\affiliation{Institut f\"ur Kernphysik, Technische Universit\"at Darmstadt, 64289 Darmstadt, Germany}

\date{\today}

%
\begin{abstract}  
We generalize the coupled-cluster (CC) approach with singles, doubles, and the non-iterative treatment of
triples termed $\Lambda$CCSD(T) to Hamiltonians containing three-body interactions.
The resulting method and the underlying CC approach with singles and doubles only (CCSD) are applied to the medium-mass closed-shell nuclei \elem{O}{16}, \elem{O}{24}, and \elem{Ca}{40}.
By comparing the results of  CCSD and $\Lambda$CCSD(T) calculations with explicit treatment of three-nucleon interactions
to those obtained using an approximate treatment in which they are included
effectively via the zero-, one-, and two-body components of the Hamiltonian in normal-ordered form,
we quantify the contributions of the residual three-body interactions neglected in the approximate treatment.
We find these residual normal-ordered three-body contributions negligible for the
$\Lambda$CCSD(T) method, although they can become significant in the lower-level CCSD approach,
particularly when the nucleon-nucleon interactions are soft.
\end{abstract}


\maketitle


%
\section{Introduction}
\label{sec1}

Chiral effective field theory (EFT)
provides a systematic link between low-energy quantum chromodynamics (QCD) and
nuclear-structure physics~\cite{Weinberg79,GaLe84,Weinberg90,Weinberg91,VanKolck99,EpHa09,MaEn11}.
In order to make accurate QCD-based predictions using {\it ab initio} many-body methods
employing Hamiltonians constructed within chiral
EFT, the inclusion of three-nucleon (3$N$) forces is inevitable~\cite{MaEn11,EpHa09},
affecting various important nuclear properties, such as binding and excitation energies~\cite{NaGu07,NoNa06,RoLa11,RoBi12,MaVa13,HeBo13,HeBi13}.
While some many-body approaches, such as the no-core shell model (NCSM)~\cite{NaVa00,NaKa00,NaOr02,RoNa07,CaMa05,NaQu09}
and its importance-truncated (IT) extension~\cite{Roth09,RoNa07,nuclei10} or 
coupled-cluster (CC) theory~\cite{Coester:1958,Coester:1960,kummel-1978,cizek1,cizek2,cizek3,cizek4,bishop-1998} truncated
at the singly and doubly excited clusters (CCSD)~\cite{ccsd,ccsd2,nuclei1,gmatrix2,nuclei3,nuclei4,nuclei5,nuclei8,HaPa08,HaPa09b,nuclei10,nuclei11,HaHj12}
have already been extended to the explicit treatment of 3$N$ interactions and were successfully applied to light and
medium-mass nuclei~\cite{HaPa07,RoLa11,RoBi12,HeBi13,BiLa13},
other approaches remain to be generalized to the explicit 3$N$ case. 
Among these are the more quantitative CC approaches, including those based on a non-iterative
treatment of the connected triply excited clusters on top of CCSD, such as CCSD(T)~\cite{ccsdpt,HaPa07},
CR-CCSD(T)~\cite{nuclei1,nuclei2,nuclei3,nuclei4,nuclei5,leszcz,ren1,ren2,irpc,PP:TCA},
CCSD(2)$_{\rm T}$~\cite{gwaltney1,gwaltney3,eomccpt,ccsdpt2},
$\Lambda$CCSD(T)~\cite{ref:26,bartlett2008a,bartlett2008b,HaPa09b,HaPa10,HeBi13,BiLa13},
and CR-CC(2,3)~\cite{crccl_jcp,crccl_cpl,crccl_jpc,crccl_ijqc2,nuclei8,nuclei10,nuclei11}, or the in-medium similarity renormalization group~\cite{TsBo11,HeBo13}.

Considering the substantial cost of  {\it ab initio} many-body computations with 3$N$ interactions,
it is important to examine how much information about the 3$N$ forces has to be included
in such calculations explicitly. 
A common practice in nuclear-structure theory is to
incorporate 3$N$ forces into the many-body considerations with the help of
effective interactions that can provide information about these forces
via suitably re-defined lower-particle terms in the Hamiltonian.
In particular, the normal-ordering two-body approximation (NO2B), where normal ordering of the Hamiltonian
becomes a formal tool to demote information about the 3$N$ interactions to  lower-particle normal-ordered terms and the residual
normal-ordered 3$N$ term is subsequently discarded, has led to promising results
in  NCSM and CCSD calculations for light and medium-mass nuclei~\cite{HaPa07,RoLa11,RoBi12,HeBi13,BiLa13}. In the case of the IT-NCSM and CCSD approach, contributions from the residual 3$N$ interactions have been shown to be small~\cite{HaPa07,RoBi12,BiLa13},
although not always negligible~\cite{RoBi12,BiLa13}. 
In many cases one needs to go beyond the CCSD level within the CC framework to obtain a highly accurate
quantitative description of several nuclear properties, including binding and excitation 
energies~\cite{nuclei1,nuclei2,nuclei3,nuclei4,nuclei5,nuclei7,nuclei8,nuclei10,nuclei11,HaPa09b,HaPa10}.
Thus, a more precise assessment of the significance of the residual 3$N$ contribution
in the normal-ordered Hamiltonian at the CC theory levels that incorporate the connected triply excited clusters
in an accurate and computationally manageable manner, such as CCSD(T), $\Lambda$CCSD(T) and CR-CC(2,3), is an important and timely objective. 
It is nowadays well established that once the connected triply excited clusters
are included in the CC calculations, the resulting energies can compete with the converged
NCSM, high-level configuration interaction (CI), or other nearly exact numerical data, which is
a consequence of the use of the exponential wave function ansatz in the CC considerations,
where various higher-order many-particle correlation effects are described via
products of low-rank excitation operators
(for the examples of the more recent nuclear-structure calculations
illustrating this statement, see Refs.~\cite{nuclei1,nuclei2,nuclei3,nuclei4,nuclei5,nuclei7,nuclei8,HaPa09b,nuclei10,nuclei11,HaHj12,HaPa07,HaPa10,HeBi13,BiLa13};
cf., also, Ref.~\cite{mihaila-1999}).
This makes the examination of the CC models that account for the connected triply excited clusters, in addition to the singly and doubly excited clusters and their products captured by CCSD, and their extensions to 3$N$ interactions even more important.

In our earlier work on CC methods with non-iterative treatment of
the connected triply excited clusters (called triples) using two-nucleon ($NN$) interactions
in the Hamiltonian, the highest theory level considered thus far was
CR-CC(2,3)~\cite{nuclei10,nuclei11}. The experience of quantum chemistry, where several
CC approximations of this type have been developed, indicates
that CR-CC(2,3) represents the most complete and most
robust form of the non-iterative triples correction to CCSD
(cf., e.g., Refs.~\cite{crccl_jcp,crccl_cpl,crccl_jpc,ge1,ge2,taube2010,jspp-chemphys2012}),
producing results that in benchmark computations
are often very close to those obtained with a full treatment of the
singly, doubly, and triply excited clusters via the iterative CCSDT approach~\cite{ccfullt,ccfullt2}, at a small fraction of the 
computing cost~\cite{crccl_jcp,crccl_cpl,jspp-chemphys2012}.
However, there also
exist other methods in this category, such as the $\Lambda$CCSD(T) approach that has been
examined in the nuclear context as well~\cite{HaPa10,HeBi13,BiLa13}, which
represent approximations to CR-CC(2,3)~\cite{crccl_jcp,crccl_cpl,crccl_jpc,jspp-chemphys2012} and  are almost as
effective  in capturing the connected triply excited clusters in closed-shell systems, while simplifying programming effort, particularly
when 3$N$ interactions need to be examined and when efficient angular-momentum-coupled codes have to be developed. Thus, although we would eventually also like to work
on an angular-momentum-coupled formulation of the
CR-CC(2,3) method for Hamiltonians including 3$N$ forces, in this first work on the examination of the role of 3$N$ interactions in the CC theory levels beyond CCSD, we focus on the simpler $\Lambda$CCSD(T) approach.
Following the considerations presented in Ref.~\cite{bartlett2008a} for the case of two-body Hamiltonians and
those presented in Refs.~\cite{crccl_jcp,crccl_cpl,crccl_ijqc2} in the more general CR-CC(2,3) context, which help us to identify additional terms in the equations due to the 3$N$ forces,
we derive the $\Lambda$CCSD(T)-style triples energy correction for three-body Hamiltonians 
which we subsequently apply to the medium-mass closed-shell nuclei \elem{O}{16}, \elem{O}{24}, and \elem{Ca}{40}.
By comparing the CCSD and $\Lambda$CCSD(T) binding energies obtained with the explicit treatment of 3$N$ interactions
with their counterparts obtained within the NO2B approximation, we quantify the contributions of the residual 3$N$ interactions that are neglected in the NO2B approximation at two different CC-theory levels, with and without the connected triply excited clusters.

%
\section{Theory}
\label{sec2}

%
\subsection{Brief synopsis of coupled-cluster theory}
\label{sec2A}

The CCSD and $\Lambda$CCSD(T) approaches examined in this study, and the CR-CC(2,3) counterpart of $\Lambda$CCSD(T) used in our considerations as well,
are examples of approximations based on the
exponential ansatz of  single-reference CC theory, in which the
ground state $|\Psi\rangle$ of an $A$-particle system is represented 
as~\cite{Coester:1958,Coester:1960,kummel-1978,cizek1,cizek2,cizek3,cizek4,bishop-1998}
\beq
|\Psi\rangle= e^{T} |\Phi\rangle ,
\label{eq:cc}
\eeq
where $|\Phi\rangle$ is the reference determinant (in the computations reported in this paper, the
Hartree-Fock state) and
\beq
T = \sum_{n=1}^{A} T_{n}
\label{eq:tdef}
\eeq
is a particle-hole
excitation operator, defined relative to the Fermi vacuum $|\Phi\rangle$ and
referred to as the cluster operator, whose many-body components
\beq
T_{n} = \left( \frac{1}{n!} \right)^{2}
\sum_{i_{1}, \ldots, i_{n} \atop a_{1}, \ldots, a_{n}}
t_{i_{1}\ldots i_{n}}^{a_{1} \ldots a_{n}} \,
a_{a_{1}}^{\dagger} \cdots a_{a_{n}}^{\dagger} a_{i_{n}} \cdots a_{i_{1}}
\label{eq:tn}
\eeq
generate the connected wave-function diagrams of $|\Psi\rangle$.
The remaining linked, but disconnected contributions to $|\Psi\rangle$ are produced through
the various product terms of the $T_{n}$ operators resulting from the use of Eqs.~(\ref{eq:cc})--(\ref{eq:tn}).
Here and elsewhere in this article,
we use the traditional notation in which
$i_{1}, \: i_{2}, \ldots$ or $i, \: j, \ldots$ denote the single-particle states (orbitals)
occupied in $|\Phi\rangle$,
$a_{1}, \: a_{2}, \ldots$ or $a, \: b, \ldots$ denote the single-particle states
unoccupied in $|\Phi\rangle$, and $p,q,\ldots \:$, $p_{1}, \: p_{2}, \ldots \:$, or
$q_{1}, \: q_{2}, \ldots$ represent generic single-particle states.

Typically, the explicit equations for the ground-state energy $E$, which can be written as
\beq
E = \ERef + \Delta E ,
\label{energy}
\eeq
where
\beq
\ERef = \langle \Phi |H | \Phi \rangle
\label{referenceenergy}
\eeq
is the independent-particle-model reference energy and
$\Delta E$ its correlation counterpart, and the cluster amplitudes
$t_{i_{1}\ldots i_{n}}^{a_{1} \ldots a_{n}}$ defining the many-body components $T_{n}$ of $T$,
are obtained by first inserting the ansatz for the wave function $|\Psi\rangle$, Eq.~(\ref{eq:cc}),
into the Schr{\" o}dinger equation, $H_{N} |\Psi\rangle = \Delta E |\Psi\rangle$, where
\beq
H_{N} = H - \ERef
\label{normalH}
\eeq
is the Hamiltonian in normal-ordered form relative to $|\Phi\rangle$.
Then, premultiplying both sides of the resulting equation on the left by $e^{-T}$ yields
the connected cluster form of the Schr{\" o}dinger equation~\cite{cizek1,cizek2},
\beq
\overline{H_{N}} |\Phi \rangle = \Delta E |\Phi \rangle ,
\label{rightcc}
\eeq
where
\beq
\overline{H_{N}}= e^{-T} \, H_{N} \,  e^{T} = (H_{N} \, e^{T})_{C}
\label{similaritycc}
\eeq
is the similarity-transformed Hamiltonian
or, equivalently, the connected product of $H_{N}$ and $e^{T}$
(designated by the subscript $C$).
Finally, both sides of Eq.~(\ref{rightcc}) are projected on the reference determinant $|\Phi\rangle$ and
the excited determinants
\beq
|\Phi_{i_{1} \ldots i_{k}}^{a_{1} \ldots a_{k}} \rangle
= a_{a_{1}}^{\dagger} \cdots a_{a_{k}}^{\dagger} a_{i_{k}} \cdots a_{i_{1}} |\Phi\rangle
\label{excited-determinant}
\eeq
that correspond to the particle-hole excitations included in $T$. The latter projections result in a nonlinear
system of explicitly connected and energy-independent equations for the cluster amplitudes
$t_{i_{1}\ldots i_{k}}^{a_{1} \ldots a_{k}}$
\cite{cizek1,cizek2,cizek3,cizek4}
(cf., e.g.,
Refs.~\cite{kummel-1978,bishop-1998,gauss,paldus-li,PP:TCA,ptcp2006,bartlett-musial2007,nucbook1,nuclei10,jspp-chemphys2012}
for review information),
\beq
\langle \Phi_{i_{1} \ldots i_{n}}^{a_{1} \ldots a_{n}} |
\overline{H_{N}} |\Phi\rangle = 0 ,
\;\; i_{1}< \cdots < i_{n}, \;
a_{1} < \cdots < a_{n},
\label{cceqs}
\eeq
where $\overline{H_{N}}$ is defined by Eq.~(\ref{similaritycc}) and $n=1,\ldots,A$, whereas the projection
of Eq.~(\ref{rightcc}) on $|\Phi\rangle$ results in the CC correlation energy formula,
\beq
\Delta E = \langle\Phi|\overline{H_{N}}|\Phi\rangle .
\label{ccenergy}
\eeq

If one is further interested in properties other than energy, which require the knowledge of the
ket state $|\Psi\rangle$ and its bra counterpart
\beq
\langle \tilde{\Psi}| = \langle \Phi | (1 + \Lambda ) e^{-T} ,
\label{bracc}
\eeq
which satisfies the biorthonormality condition
$\langle\tilde{\Psi} | \Psi \rangle = 1$, and where
\beq
\Lambda = \sum_{n=1}^{A} \Lambda_{n} ,
\label{eq:lambdadef}
\eeq
with
\beq
\Lambda_{n} = \left( \frac{1}{n!} \right)^{2}
\sum_{i_{1}, \ldots, i_{n} \atop a_{1}, \ldots, a_{n}}
\lambda_{a_{1}\ldots a_{n}}^{i_{1} \ldots i_{n}} \,
a_{i_{1}}^{\dagger} \cdots a_{i_{n}}^{\dagger} a_{a_{n}} \cdots a_{a_{1}} ,
\label{eq:lambdan}
\eeq
is the hole-particle deexcitation operator generating $\langle\tilde{\Psi} |$,
we also have to solve the linear system of the so-called
$\Lambda$ equations~\cite{ccgrad,ccsdgrad,eomcc3,pp_rjb,gauss,ptcp2006,bartlett-musial2007,nucbook1,nuclei10,jspp-chemphys2012},
\begin{eqnarray}
\langle \Phi | (1 + \Lambda) \: \overline{H_{N}} | \Phi_{i_{1} \ldots i_{n}}^{a_{1} \ldots a_{n}} \rangle & = & 
\Delta E \: \lambda_{a_{1}\ldots a_{n}}^{i_{1} \ldots i_{n}} ,
\nonumber
\\
&&
\!\!\!\!\!
\!\!\!\!\!
\!\!\!\!\!
i_{1}< \cdots < i_{n}, \;
a_{1} < \cdots < a_{n},
\label{lambdacceqs}
\end{eqnarray}
obtained by substituting Eq.~(\ref{bracc}) into the adjoint form of the
Schr{\" o}dinger equation, $\langle \tilde{\Psi} | H_{N} = \Delta E \langle \tilde{\Psi} |$.
System~(\ref{lambdacceqs}) can be further simplified into the energy-independent form
\begin{eqnarray}
\langle \Phi | (1 + \Lambda) \: (\overline{H_{N}})_{\rm open} | \Phi_{i_{1} \ldots i_{n}}^{a_{1} \ldots a_{n}} \rangle & = & 0 ,
\nonumber
\\
&&
\!\!\!\!\!
\!\!\!\!\!
\!\!\!\!\!
\!\!\!\!\!
\!\!\!\!\!
i_{1}< \cdots < i_{n}, \;
a_{1} < \cdots < a_{n},
\label{lambdacceqs-better}
\end{eqnarray}
where
\beq
(\overline{H_{N}})_{\rm open} = \overline{H_{N}} - (\overline{H_{N}})_{\rm closed}
= \overline{H_{N}} - \Delta E
\label{eq14}
\eeq
is the open part of $\overline{H_{N}}$,
defined by the diagrams of $\overline{H_{N}}$ that have external Fermion lines.
Clearly, the only diagrams of $\overline{H_{N}}$ that enter the CC system given by Eq.~(\ref{cceqs}) are
the diagrams of $(\overline{H_{N}})_{\rm open}$, whereas the only diagrams that
contribute to $\Delta E$, Eq.~(\ref{ccenergy}), are the vacuum (or closed) diagrams
that have no external lines. We discuss the $\Lambda$ or left-eigenstate CC equations,
Eq.~(\ref{lambdacceqs}) or (\ref{lambdacceqs-better}),
for the deexcitation amplitudes $\lambda_{a_{1}\ldots a_{n}}^{i_{1} \ldots i_{n}}$
here, since they are one of the key
ingredients of $\Lambda$CCSD(T)  and the related CR-CC(2,3) considerations below.
It is worth pointing out, though, that by
examining these equations in the
context of the $\Lambda$CCSD(T)/CR-CC(2,3) considerations for three-body Hamiltonians,
we are at the same time helping future developments in the area of CC computations of nuclear properties
other than binding energy, extending the relevant formal considerations
to the case of 3$N$ interactions. For example,  the $\Lambda$ operator obtained by solving Eq.~(\ref{lambdacceqs-better}) can be used to determine
the CC one-body reduced density matrices,
\beq
\gamma_{p}^{q}
\equiv \langle\tilde{\Psi} | ( a_{p}^{\dagger} a_{q} ) | \Psi \rangle
= \langle \Phi | (1+\Lambda) \overline{(a_{p}^{\dagger} a_{q})} | \Phi \rangle ,
\label{cc-one-rdm}
\eeq
where we define $\overline{(a_{p}^{\dagger} a_{q})}$ as
\beq
\overline{(a_{p}^{\dagger} a_{q})} = e^{-T} \, (a_{p}^{\dagger} a_{q}) \, e^{T}
= [(a_{p}^{\dagger} a_{q}) \, e^{T}]_{C} ,
\label{cc-adagger-a}
\eeq
and determine expectation values of one-body operators in the usual manner as
\beq
  \langle \tilde{\Psi} | \Theta | \Psi \rangle
  = \sum_{p,q} \theta_{q}^{p} \, \gamma_{p}^{q} \equiv \theta_{q}^{p} \, \gamma_{p}^{q} ,
\label{cc-property}
\eeq
where $\Theta = \sum_{p,q} \theta_{q}^{p} a_{p}^{\dagger} a_{q}$ is a
one-body property operator of interest. In writing Eq.~(\ref{cc-property}), the
Einstein summation convention over repeated upper and lower indices in product expressions of matrix elements has been
assumed. We will exploit this convention throughout the rest of this article.

The above is the exact CC theory, which is equivalent to the exact diagonalization
of the Hamiltonian within the full CI approach and is, for practical reasons, limited to
small many-body problems. Thus, in all practical applications of CC theory, one truncates the many-body expansion for $T$, Eq.~(\ref{eq:tdef}), at some,
preferably low, $m$-particle--$m$-hole excitation level $T_{m}$. In this study, we focus
on the CCSD approach
in which $T$ is truncated at the doubly excited clusters $T_{2}$, and the
$\Lambda$CCSD(T) and CR-CC(2,3) methods, which allow one to correct the CCSD
energy for the dominant effects due to the triply excited clusters $T_{3}$ in a computationally
feasible manner, avoiding the prohibitively expensive steps of full CCSDT, in which one has to solve for
$T_{1}$, $T_{2}$ and $T_{3}$ in an iterative fashion.
The final form of the CC amplitude and energy equations also depends on the Hamiltonian used in the
calculations, since the length of the many-body expansion of the
resulting similarity-transformed Hamiltonian $\overline{H_{N}}$, Eq.~(\ref{similaritycc}),
which can also be written as
\begin{eqnarray}
\overline{H_{N}} & = &
H_N + \sum_{n=1}^{2k_{\rm max}} \ \frac{1}{n!} \ \underbrace{
        \Big[ \dots \Big[
}_{n \ \text{times}}
        H_N,
\underbrace{
T \Big]  , \dots , T \Big]
}_{n \ \text{times}}
\nonumber
\\
& = &
\sum_{n=0}^{2k_{\rm max}} \ \frac{1}{n!} (H_{N} T^{n})_{C} ,
\label{CHB}
\end{eqnarray}
depends on $k_{\rm max}$,
where $k_{\rm max}$ is the highest many-body rank of the interactions in $H_{N}$ or $H$ ($k_{\rm max} = 2$ for $2N$ interactions,
$k_{\rm max} = 3$ for $3N$ interaction terms, etc.). In this article 
we focus on 
the $k_{\rm max} = 3$ case, emphasizing
the differences between the more familiar CCSD and $\Lambda$CCSD(T) equations for two-body Hamiltonians,
which can be found, in the most compact, factorized form using recursively generated intermediates,
in Refs.~\cite{ccsd,nuclei6,gmatrix2,ccsdgrad} for CCSD and
\cite{bartlett2008a} for $\Lambda$CCSD(T), and their extensions to the three-body case.
The key ingredients of the CCSD and $\Lambda$CCSD(T)-type approaches for 
3$N$ interactions in the Hamiltonian are discussed in the next two subsections. We begin with the Hamiltonian.

%
%
\subsection{Normal-ordered form of the Hamiltonian with three-body interactions and the NO2B approximation}
\label{sec2B}

As shown in the previous subsection, the single-reference CC equations for the cluster
amplitudes $t_{i_{1}\ldots i_{n}}^{a_{1} \ldots a_{n}}$ defining $T$, their deexcitation counterparts
$\lambda_{a_{1}\ldots a_{n}}^{i_{1} \ldots i_{n}}$ defining $\Lambda$,
 and the correlation energy $\Delta E$ 
can be conveniently expressed in terms of the Hamiltonian in  normal-ordered form
relative to the Fermi vacuum $|\Phi\rangle$, transformed with $e^{T}$,
as in Eqs.~(\ref{similaritycc}) and~(\ref{CHB}). For 
Hamiltonians with up to three-body interactions,
\beq
H = H_{1} + H_{2} + H_{3} ,
\label{hamiltonian}
\eeq
where
\beq
H_{n} = \left( \frac{1}{n!} \right)^{2}
\sum_{p_{1}, \ldots, p_{n} \atop q_{1}, \ldots, q_{n}}
h_{q_{1}\ldots q_{n}}^{p_{1} \ldots p_{n}} \,
a_{p_{1}}^{\dagger} \cdots a_{p_{n}}^{\dagger} a_{q_{n}} \cdots a_{q_{1}}
\label{eq:Hn}
\eeq
is the $n$-body contribution to $H$, and the normal-ordered Hamiltonian $H_{N}$, Eq.~(\ref{normalH}),
which provides information about the many-particle correlation effects beyond the mean-field level
represented by $|\Phi\rangle$,
can be represented in the form
\beq
H_{N} = F_{N} + V_{N} + W_{N} .
\label{normal-form}
\eeq
The one-, two-, and three-body components $F_N$, $V_N$ and $W_N$ in Eq.~(\ref{normal-form}) are defined as
\begin{eqnarray}
F_{N} &=& \sum_{p,q} f_{q}^{p} \, N[a_{p}^{\dagger} a_{q}] ,
\label{FN} 
\end{eqnarray}
\begin{eqnarray}
V_{N} &=& \qua \sum_{p,q,r,s} v_{rs}^{pq} \, N[a_{p}^{\dagger} a_{q}^{\dagger} a_{s} a_{r}] ,
\label{VN} 
\end{eqnarray}
and
\begin{eqnarray}
W_{N} &=& \tsix \sum_{p,q,r,s,t,u} w_{stu}^{pqr} \, N[a_{p}^{\dagger} a_{q}^{\dagger} a_{r}^{\dagger} a_{u} a_{t} a_{s}] ,
\label{WN}
\end{eqnarray}
where $N[\ldots]$ designates normal ordering relative to $|\Phi\rangle$ and the matrix elements $f_{q}^{p}$, $v_{rs}^{pq}$ and $w_{stu}^{pqr}$ are given by
\begin{eqnarray}
f_{q}^{p} &=& h_{q}^{p} + \sum_{i} h_{qi}^{pi} + \half \sum_{i,j} h_{qij}^{pij} ,
\label{fpq} 
\end{eqnarray}
\begin{eqnarray}
v_{rs}^{pq} &=& h_{rs}^{pq} + \qua \sum_{i}  h_{rsi}^{pqi} ,
\label{vpqrs} 
\end{eqnarray}
and
\begin{eqnarray}
w_{stu}^{pqr} &=& h_{stu}^{pqr} ,
\label{wpqrstu}
\end{eqnarray}
respectively.
The corresponding reference energy $\ERef$, Eq.~(\ref{referenceenergy}), which one needs to
add to the correlation energy $\Delta E$ to obtain the total ground-state energy $E$, is calculated via
\beq
\ERef = \sum_{i} h_{i}^{i} + \half \sum_{i,j} h_{ij}^{ij} + \six \sum_{i,j,k} h_{ijk}^{ijk} .
\label{ref-energy}
\eeq

When the Hamiltonian is used in the normal-ordered form, information about the three-body interaction in $H$
enters in two fundamentally different ways: effectively, via the reference energy $\ERef$, Eq.~(\ref{ref-energy}), and the normal-ordered one- and two-body matrix elements
$f_{q}^{p}$ and $v_{rs}^{pq}$, Eqs.~(\ref{fpq}) and~(\ref{vpqrs}), which define the $F_{N}$ and $V_{N}$ components
of $H_{N}$, and explicitly, via the genuinely three-body residual term $W_{N}$, Eq.~(\ref{WN}),
which captures those 3$N$ contributions to the Hamiltonian that cannot be demoted to the lower-rank
$F_{N}$ and $V_{N}$ operators or the reference energy $\ERef$. Considering the fact that the
$F_{N}$ and $V_{N}$ components of $H_{N}$ combined with the reference energy $\ERef$
contain the complete information about pairwise interactions and much of the information about the 3$N$ forces,
it is reasonable to consider the NO2B approximation, discussed in
Refs.~\cite{HaPa07,RoBi12,HeBi13,BiLa13}, in which the three-body residual term $W_{N}$ is neglected in $H_{N}$. The main goal of this study is to compare the CCSD and $\Lambda$CCSD(T)-type results obtained with a full representation of the normal-ordered Hamiltonian $H_{N}$ in which the residual three-body term $W_{N}$ is retained in the
calculations, with their counterparts obtained using the truncated form of $H_{N}$
that defines the NO2B approximation, in which Eq.~(\ref{normal-form}) is replaced by the simplified expression
\beq
H_{N,{\rm 2B}} = F_{N} + V_{N}
\label{HN-NO2B}
\eeq
containing only the one- and two-body components of $H_{N}$ defined
by Eqs.~(\ref{FN})--(\ref{VN}) and~(\ref{fpq})--(\ref{vpqrs}).

The NO2B approximation offers several advantages over the
full treatment of 3$N$ forces. First of all, it allows 
to reuse the conventional CC equations derived for two-body Hamiltonians, which one can find for CCSD
in Refs.~\cite{ccsd,nuclei6,gmatrix2,ccsdgrad} and for $\Lambda$CCSD(T) in Ref.~\cite{bartlett2008a},
by replacing the $f_{q}^{p}$ and $v_{rs}^{pq}$ matrix elements in these equations
with their values determined using Eqs.~(\ref{fpq}) and~(\ref{vpqrs}). Clearly,
the three-body interactions are not ignored when the NO2B approximation is
invoked, since the reference energy $\ERef$, Eq.~(\ref{ref-energy}), the one-body operator $F_{N}$, defined by Eqs.~(\ref{FN}) and~(\ref{fpq}), and the two-body
operator $V_{N}$, defined by Eqs.~(\ref{VN}) and~(\ref{vpqrs}), contain information about the
3$N$ forces in the form of the integrated $\tfrac{1}{6}\sum_{i,j,k}h^{ijk}_{ijk}$, $\half \sum_{i,j} h_{qij}^{pij}$
and $\qua \sum_{i} h_{rsi}^{pqi}$ contributions to $\ERef$,
$f_{q}^{p}$ and $v_{rs}^{pq}$. Secondly, the NO2B approximation
leads to major savings in the computational effort, since the most expensive terms in the CC equations that are generated
by the three-body residual interaction $W_{N}$ are disregarded when one uses Eq.~(\ref{HN-NO2B}) instead of
Eq.~(\ref{normal-form}). Our objective is to examine if neglecting these residual terms, particularly
at the more quantitative $\Lambda$CCSD(T) level, does not result in a substantial loss of accuracy in the description of the 3$N$
contributions to the resulting binding energies.

The above discussion implies that in order to compare the CCSD and $\Lambda$CCSD(T) energies
corresponding to the full treatment of 3$N$ forces with their counterparts obtained using
the NO2B approximation, as defined by Eq.~(\ref{HN-NO2B}), one has to augment the existing
CCSD and $\Lambda$CCSD(T) equations derived for Hamiltonians with up to
two-body components in $H_{N}$, reported, for example,
in Refs.~\cite{ccsd,nuclei6,gmatrix2,ccsdgrad,bartlett2008a},
by terms generated by the residual $W_{N}$ interaction,
while adjusting matrix elements of the $F_{N}$ and $V_{N}$ operators in the resulting
equations through the use of
Eqs.~(\ref{fpq}) and~(\ref{vpqrs}). This has been done for the CCSD case in
Ref.~\cite{HaPa07}, but none of the earlier nuclear CC works have dealt with the
explicit and complete incorporation of 3$N$ interactions in modern post-CCSD considerations.
The present study addresses this concern by extending the considerations
reported in Ref.~\cite{HaPa07} to the triples energy correction of $\Lambda$CCSD(T) and, also,
the $\Lambda$CCSD equations, which one has to solve prior to the determination of
$\Lambda$CCSD(T)- or CR-CC-type corrections. Since, as discussed in Sec.~\ref{sec2A},
the CC amplitude and energy equations and their
left-eigenstate $\Lambda$ counterparts rely on the similarity-transformed form of
$H_{N}$, designated by $\overline{H_{N}}$, Eq.~(\ref{similaritycc}), the most convenient
way to incorporate the additional terms due to the presence of $W_N$ into the CC considerations is by partitioning $\overline{H_{N}}$ as 
\beq
\overline{H_{N}} = e^{-T} (H_{N,{\rm 2B}} + W_{N}) \, e^{T} = \overline{H_{N,{\rm 2B}}} + \overline{W_{N}} ,
\label{partitioning-of-HN}
\eeq
where
\beq
\overline{H_{N,{\rm 2B}}} = e^{-T} H_{N,{\rm 2B}} \, e^{T} = (H_{N,{\rm 2B}} \, e^{T})_{C}
\label{sim-HN-NO2B}
\eeq
is the similarity-transformed form of $H_{N,{\rm 2B}}$ and
\beq
\overline{W_{N}} = e^{-T} W_{N}  \, e^{T} = (W_{N}  \, e^{T})_{C}
\label{sim-WN}
\eeq
is the similarity-transformed form of $W_{N}$.
In this way, we can split the CC equations Eqs.~(\ref{cceqs}),~(\ref{ccenergy}) and~(\ref{lambdacceqs-better})
into the NO2B contributions expressed in terms of $\overline{H_{N,{\rm 2B}}}$, which, with the exception
of the $f_{q}^{p}$ and $v_{rs}^{pq}$ matrix elements that define $F_{N}$ and $V_{N}$, have the same algebraic
structure as the standard CC equations derived for two-body Hamiltonians, and the $W_{N}$-containing terms that provide the rest
of the information about 3$N$ contributions neglected by the NO2B approximation.

The partitioning of $\overline{H_{N}}$ represented by Eqs.~(\ref{partitioning-of-HN})--(\ref{sim-WN})
reflects the obvious fact that the normal-ordered form of the Hamiltonian including
three-body interactions, Eq.~(\ref{normal-form}), is a sum of the NO2B component
$H_{N,{\rm 2B}}$, Eq.~(\ref{HN-NO2B}), and the three-body residual $W_{N}$ term. 

As implied by Eq.~(\ref{CHB}), $\overline{H_{N,{\rm 2B}}}$
terminates at the quadruply nested commutators or terms that contain the fourth power of $T$, since one can
connect up to four vertices representing $T$ operators to the diagrams of $H_{N,{\rm 2B}}$. Similarly, $\overline{W_{N}}$
terminates at the $T^{6}$ terms, since the diagram representing $W_{N}$ has six external lines. 
As a result, the complete many-body expansions of $\overline{H_{N,{\rm 2B}}}$ and
$\overline{W_{N}}$, i.e.,
\beq
\overline{H_{N,{\rm 2B}}} = \sum_{n} \overline{H}_{n,{\rm 2B}} ,
\label{HN-NO2B-many-body}
\eeq
where
\begin{eqnarray}
\overline{H}_{n,{\rm 2B}} & = &
\left( \frac{1}{n!} \right)^{2}
\sum_{p_{1}, \ldots, p_{n} \atop q_{1}, \ldots, q_{n}}
\helem{p_{1} \ldots p_{n}}{q_{1} \ldots q_{n}}
\nonumber
\\
&& \times \:
 a_{p_{1}}^{\dagger} \cdots a_{p_{n}}^{\dagger} a_{q_{n}} \cdots a_{q_{1}} ,
\label{HNO2B-n}
\end{eqnarray}
and
\beq
\overline{W_{N}} = \sum_{n} \overline{W}_{n} ,
\label{WN-many-body}
\eeq
where
\beq
\overline{W}_{n} =
\left( \frac{1}{n!} \right)^{2}
\sum_{p_{1}, \ldots, p_{n} \atop q_{1}, \ldots, q_{n}}
\welem{p_{1}\ldots p_{n}}{q_{1} \ldots q_{n}}
\, a_{p_{1}}^{\dagger} \cdots a_{p_{n}}^{\dagger} a_{q_{n}} \cdots a_{q_{1}} ,
\label{WN-n}
\eeq
respectively, are quite complex, even at the lower levels of CC theory, such as CCSD, where $T$ is truncated at $T_{2}$.
Indeed, it is easy to demonstrate that when the cluster operator $T$ is truncated at the doubly excited $T_{2}$ component,
the resulting $\overline{H_{N,{\rm 2B}}}$ operator contains up to six-body terms. The
corresponding operator $\overline{W_{N}}$ is even more complex, containing up to nine-body terms.
Fortunately, as shown in the next subsection, by the virtue of the projections on the subsets of
determinants that enter the CCSD and $\Lambda$CCSD(T) considerations, the
final amplitude and energy equations used in the CCSD and $\Lambda$CCSD(T) calculations do not utilize all of
the many-body components of $\overline{H_{N,{\rm 2B}}}$ and $\overline{W_{N}}$. For example,
the highest many-body components of $\overline{H_{N,{\rm 2B}}}$ and $\overline{W_{N}}$ that have to be considered
in CCSD and $\Lambda$CCSD(T) calculations are selected types of three-body ($\overline{H_{N,{\rm 2B}}}$)
or four-body ($\overline{W_{N}}$) terms, which greatly simplifies these calculations.
The CCSD and $\Lambda$CCSD(T) equations, with emphasis on the additional terms beyond the NO2B approximation,
are discussed next.

%
\subsection{The CCSD and \protect\mbox{\boldmath$\Lambda$}CCSD(T) approaches for Hamiltonians with three-body interactions}
\label{sec2C}

As mentioned in the introduction, the residual 3$N$ interaction, represented by
the $W_{N}$ component of the normal-ordered Hamiltonian $H_{N}$, although generally 
small~\cite{HaPa07,RoBi12,BiLa13}, may not always be negligible, particularly when the basic CC theory level represented by the CCSD approach 
is considered~\cite{RoBi12,BiLa13}.
Considering the fact that one has to go beyond the CCSD level within the CC framework to obtain a more
quantitative description of nuclear 
properties~\cite{nuclei1,nuclei2,nuclei3,nuclei4,nuclei5,nuclei7,nuclei8,nuclei10,nuclei11,HaPa09b,HaPa10,RoBi12,HeBi13,BiLa13},
it is imperative to
investigate how significant the incorporation of the residual three-body interactions in the Hamiltonian is
when the connected triply excited ($T_{3}$) clusters are included in the calculations, in addition
to the singly and doubly excited clusters, $T_{1}$ and $T_{2}$, included in CCSD.
Ideally, one would prefer to examine this issue using the full CCSDT approach, in which one solves the
system~(\ref{cceqs}) of coupled nonlinear equations for the
$T_{1}$, $T_{2}$, and $T_{3}$ cluster components in an iterative manner.
Unfortunately,
the full CCSDT treatment is prohibitively expensive and thus limited to small many-body problems,
even at the level of pairwise interactions. When
the residual 3$N$ interactions are included in the CC considerations, the situation becomes even worse.
For this reason we resort to the approximate
treatment of the $T_{3}$ clusters via the non-iterative energy correction added to the CCSD energy
defining the $\Lambda$CCSD(T) approach, which is capable of capturing
the leading $T_{3}$ effects at the small fraction of the cost of the full CCSDT computations. A few remarks about the
closely related CR-CC(2,3) method, which contains $\Lambda$CCSD(T) as the leading approximation
and which also captures the $T_{3}$ effects,
will be given as well, since the CR-CC(2,3) expressions provide a transparent and pedagogical
mechanism for identifying terms in the $\Lambda$CCSD(T) equations that result from adding the
3$N$ interactions to the Hamiltonian. 
Considering the relatively low computational cost
of the $\Lambda$CCSD(T) approach while providing information about the
$T_{3}$ clusters, we can for medium-mass nuclei compare the results of the CC calculations describing the $T_{1}$, $T_{2}$, and $T_{3}$ effects
using the complete representation of the three-body Hamiltonian including the
residual $W_{N}$ term with their counterparts relying on the NO2B truncation of $H_{N}$.

The determination of the $\Lambda$CCSD(T) (or CR-CC(2,3)) energy, which has the general form
\beq
E = E^{\rm (CCSD)} + \delta E^{\rm (T)} ,
\label{t3-corrected-energy}
\eeq
where
\beq
E^{\rm (CCSD)} = \ERef + \Delta E^{\rm (CCSD)}
\label{ccsd-total-energy}
\eeq
is the total CCSD energy and $\delta E^{\rm (T)}$ the energy correction due to the connected $T_{3}$ clusters,
consists of four steps: first, as in all many-body computations, we generate the appropriate
single-particle basis, which in our case will be obtained from Hartree-Fock calculations. In the next
two steps, which we discuss in Sec.~\ref{sec2C-1}, we solve
the CCSD equations and their left-eigenstate $\Lambda$ counterparts, and determine the
CCSD correlation energy $\Delta E^{\rm (CCSD)}$. The $\delta E^{\rm (T)}$ correction,
discussed in Sec.~\ref{sec2C-2}, is calculated in the fourth step
using the information resulting from the CCSD and $\Lambda$CCSD calculations.

%
\subsubsection{The CCSD and left-eigenstate CCSD equations for three-body Hamiltonians}
\label{sec2C-1}

We begin our considerations with the key elements of the CCSD approach, where the cluster operator $T$
defining the ground-state wave function $|\Psi\rangle$ using Eq.~(\ref{eq:cc}) is truncated at the
doubly excited clusters, so that (cf. Eqs.~(\ref{eq:tdef}) and~(\ref{eq:tn}))
\beq
T \approx T^{\rm (CCSD)} = T_{1} + T_{2} ,
\label{tccsd}
\eeq
with
\beq
T_{1}
= \sum_{i,a} t_{i}^{a} \, a_{a}^{\dagger}a_{i} = \sum_{i,a} t_{i}^{a} \,  N[a_{a}^{\dagger}a_{i}]
\label{eq:t1}
\eeq
and
\beq
T_{2}
= \qua \sum_{i,j,a,b} t_{ij}^{ab} \, a_{a}^{\dagger}a_{b}^{\dagger}a_{j}a_{i}
= \qua \sum_{i,j,a,b} t_{ij}^{ab} \, N[a_{a}^{\dagger} a_{i} a_{b}^{\dagger} a_{j}] ,
\label{eq:t2}
\eeq
and the left-eigenstate counterpart of CCSD, where the deexcitation operator $\Lambda$
defining the bra ground state $\langle \tilde{\Psi}|$, Eq.~(\ref{bracc}), is
approximated using the expression (cf. Eqs.~(\ref{eq:lambdadef}) and~(\ref{eq:lambdan}))
\beq
\Lambda \approx \Lambda^{\rm (CCSD)} = \Lambda_{1} + \Lambda_{2} ,
\label{lamccsd}
\eeq
with
\beq
\Lambda_{1}
= \sum_{i,a} \lambda_{a}^{i} \, a_{i}^{\dagger}a_{a}
= \sum_{i,a} \lambda_{a}^{i} \, N[a_{i}^{\dagger}a_{a}]
\label{eq:lam1}
\eeq
and
\beq
\Lambda_{2}
= \qua \sum_{i,j,a,b} \lambda_{ab}^{ij} \, a_{i}^{\dagger}a_{j}^{\dagger}a_{b}a_{a}
= \qua \sum_{i,j,a,b} \lambda_{ab}^{ij} \, N[a_{i}^{\dagger} a_{a} a_{j}^{\dagger} a_{b}] .
\label{eq:lam2}
\eeq
In addition to being useful in their own right,
the CCSD and left-eigenstate CCSD calculations provide
the singly and doubly excited cluster amplitudes, $t_{i}^{a}$ and $t_{ij}^{ab}$,
and their deexcitation $\lambda_{a}^{i}$ and $\lambda_{ab}^{ij}$ analogs,
which are needed to construct the non-iterative corrections to the CCSD energy 
via the $\Lambda$CCSD(T), CR-CC(2,3), and similar techniques.
The CCSD equations for three-body Hamiltonians have been discussed in Ref.~\cite{HaPa07}, but
their left-eigenstate $\Lambda$CCSD analogs have not been examined so far. Since the regular CCSD and $\Lambda$CCSD
considerations cannot be separated out, we first summarize the CCSD amplitude and energy equations
for the case of 3$N$ interactions.

The CCSD equations are obtained by replacing $T$ in Eqs.~(\ref{cceqs}) and~(\ref{ccenergy}) by
$T^{\rm (CCSD)}$, and by limiting the projections on the excited determinants
$|\Phi_{i_{1} \ldots i_{n}}^{a_{1} \ldots a_{n}} \rangle$ in Eq.~(\ref{cceqs}) to those that correspond to
the singly and doubly excited cluster amplitudes $t_{i}^{a}$ and $t_{ij}^{ab}$ we want to determine,
so that the number of equations matches the number of 
unknowns~\cite{ccsd,ccsd2,ccsdfritz,osaccsd,nuclei1,gmatrix2,%
nuclei2,nuclei3,nuclei4,nuclei5,nuclei7,nuclei8,HaPa08,HaPa09b,nuclei10,nuclei11,HaHj12}.
Assuming that the Hamiltonian of interest contains three-body interactions, we obtain the system of
equations for $t_{i}^{a}$ and $t_{ij}^{ab}$
\cite{HaPa07}
\beq
\langle\Phi_{i}^{a}| {\overline{H_{N}}}^{\rm (CCSD)}|\Phi\rangle = \Theta_{i}^{a}({\rm 2B}) + \Theta_{i}^{a}(W_{N})
= 0 ,
\label{ccsd:eqt1}
\eeq
\beq
\langle\Phi_{ij}^{ab}|\overline{H_{N}}^{\rm (CCSD)}|\Phi\rangle = \Theta_{ij}^{ab}({\rm 2B}) + \Theta_{ij}^{ab}(W_{N})
= 0,
\label{ccsd:eqt2}
\eeq
where
\beq
\overline{H_{N}}^{\rm (CCSD)} = e^{-T_{1}-T_{2}} \, H_{N} \, e^{T_{1}+T_{2}}
= (H_{N} \, e^{T_{1}+T_{2}})_{C}
\label{hbarccsd}
\eeq
is the similarity-transformed Hamiltonian of CCSD and $|\Phi_{i}^{a}\rangle$ and
$|\Phi_{ij}^{ab}\rangle$ are the singly and doubly excited determinants relative to $|\Phi\rangle$. The
$\Theta_{i}^{a}({\rm 2B})$, $\Theta_{i}^{a}(W_{N})$,
$\Theta_{ij}^{ab}({\rm 2B})$, and $\Theta_{ij}^{ab}(W_{N})$
terms entering Eqs.~(\ref{ccsd:eqt1}) and~(\ref{ccsd:eqt2}) are defined as
\beq
\Theta_{i}^{a}({\rm 2B}) = \langle\Phi_{i}^{a}| \overline{H_{N,{\rm 2B}}}^{\rm (CCSD)} |\Phi\rangle ,
\label{theta1-NO2B}
\eeq
\beq
\Theta_{i}^{a}(W_{N}) = \langle\Phi_{i}^{a}| \overline{W_{N}}^{\rm (CCSD)} |\Phi\rangle ,
\label{theta1-WN}
\eeq
\beq
\Theta_{ij}^{ab}({\rm 2B}) = \langle\Phi_{ij}^{ab}| \overline{H_{N,{\rm 2B}}}^{\rm (CCSD)} |\Phi\rangle ,
\label{theta2-NO2B}
\eeq
and
\beq
\Theta_{ij}^{ab}(W_{N}) = \langle\Phi_{ij}^{ab}| \overline{W_{N}}^{\rm (CCSD)} |\Phi\rangle .
\label{theta2-WN}
\eeq
The operators $\overline{H_{N,{\rm 2B}}}^{\rm (CCSD)}$ and $\overline{W_{N}}^{\rm (CCSD)}$ appearing in Eqs.~(\ref{theta1-NO2B})--(\ref{theta2-WN}) are defined as
\beq
\overline{H_{N,{\rm 2B}}}^{\rm (CCSD)} = e^{-T_{1}-T_{2}} \, H_{N,{\rm 2B}} \, e^{T_{1}+T_{2}}
= (H_{N,{\rm 2B}} \, e^{T_{1}+T_{2}})_{C}
\label{hbar-NO2B-ccsd}
\eeq
and
\beq
\overline{W_{N}}^{\rm (CCSD)} = e^{-T_{1}-T_{2}} \, W_{N} \,  e^{T_{1}+T_{2}}
= (W_{N} \, e^{T_{1}+T_{2}})_{C}  ,
\label{sim-WN-ccsd}
\eeq
and represent the similarity-transformed forms of the $H_{N,{\rm 2B}}$ and $W_{N}$ operators,
Eqs.~(\ref{sim-HN-NO2B}) and~(\ref{sim-WN}), adapted to the CCSD case, which obviously add up to
$\overline{H_{N}}^{\rm (CCSD)}$,
\beq
\overline{H_{N,{\rm 2B}}}^{\rm (CCSD)} + \overline{W_{N}}^{\rm (CCSD)} = \overline{H_{N}}^{\rm (CCSD)} .
\label{hbarccsd-sum}
\eeq
From the above definitions it is apparent that $\Theta_{i}^{a}(W_{N})$ and $\Theta_{ij}^{ab}(W_{N})$,
which originate from $W_N$, contribute only  when the residual 3$N$ interaction is included in the calculations, whereas the NO2B contributions $\Theta_{i}^{a}({\rm 2B})$ and $\Theta_{ij}^{ab}({\rm 2B})$ are present in any case.
As in the most common case of two-body Hamiltonians (see, e.g., 
Refs.~\cite{ccsd,ccsd2,ccsdfritz,osaccsd,nucbook1,nuclei10}), it is easy to demonstrate, using
Eq.~(\ref{CHB}) for $k_{\rm max} = 2$ and the above definitions of
$\Theta_{i}^{a}({\rm 2B})$ and $\Theta_{ij}^{ab}({\rm 2B})$, that 
the NO2B contributions to the CCSD amplitude equations do not contain higher--than--quartic terms in $T$, i.e.,
\begin{eqnarray}
\Theta_{i}^{a}({\rm 2B}) & = & \langle \Phi_{i}^{a} | \,
[H_{N,{\rm 2B}} \, (1 + T_{1} + T_{2} + \half T_{1}^{2}
\nonumber \\
&& + T_{1} T_{2}
+ \six T_{1}^{3} )]_{C}
|\Phi \rangle
\label{momccsd1-NO2B}
\end{eqnarray}
and
\begin{eqnarray}
\Theta_{ij}^{ab}({\rm 2B}) & = & \langle \Phi_{ij}^{ab} | \,
[H_{N,{\rm 2B}} \, (1 + T_{1} + T_{2} + \half T_{1}^{2}
\nonumber \\
&& + T_{1} T_{2}
+ \six T_{1}^{3}
+ \half T_{2}^{2}
+ \half T_{1}^{2} T_{2}
\nonumber \\
&& + \tfour T_{1}^{4} )]_{C}
|\Phi \rangle .
\label{momccsd2-NO2B}
\end{eqnarray}
For the $\Theta_{i}^{a}(W_{N})$ and $\Theta_{ij}^{ab}(W_{N})$ contributions to the CCSD amplitude equations due to
the residual three-body interaction term $W_{N}$, we can write~\cite{HaPa07}
\begin{eqnarray}
\Theta_{i}^{a}(W_{N}) & = & \langle \Phi_{i}^{a} | \,
[W_{N} \, (T_{2} + \half T_{1}^{2} + T_{1} T_{2} + \six T_{1}^{3}
\nonumber \\
&& + \half T_{2}^{2} + \half T_{1}^{2} T_{2} + \tfour T_{1}^{4} )]_{C}
|\Phi \rangle
\label{momccsd1-WN}
\end{eqnarray}
and
\begin{eqnarray}
\Theta_{ij}^{ab}(W_{N}) & = & \langle \Phi_{ij}^{ab} | \,
[W_{N} \, (T_{1} + T_{2} + \half T_{1}^{2} + T_{1} T_{2} + \six T_{1}^{3}
\nonumber \\
&& + \half T_{2}^{2} + \half T_{1}^{2} T_{2} + \tfour T_{1}^{4} + \half  T_{1} T_{2}^{2} 
\nonumber \\
&& + \six T_{1}^{3} T_{2} + \onetwenty T_{1}^{5} )]_{C}
|\Phi \rangle ,
\label{momccsd2-WN}
\end{eqnarray}
respectively,
i.e., the highest power of $T$ that needs to be considered is 5, not 6, as Eq.~(\ref{CHB})
 for the $k_{\rm max} = 3$ case would imply, since diagrams of the
$(W_{N} T^{6})_{C}$ type entering $\overline{W_{N}}$ have more than four external lines and, as such,
cannot produce non-zero expressions when projected on $|\Phi_{i}^{a}\rangle$ and $|\Phi_{ij}^{ab}\rangle$.

The detailed $m$-scheme-style expressions for the NO2B-type $\Theta_{i}^{a}({\rm 2B})$ and $\Theta_{ij}^{ab}({\rm 2B})$
contributions to the CCSD amplitude equations, in terms of the one- and two-body matrix elements
of the normal-ordered Hamiltonian $f_{q}^{p}$ and $v_{rs}^{pq}$, and the singly and doubly
excited cluster amplitudes $t_{i}^{a}$ and $t_{ij}^{ab}$, which lead to efficient computer
codes through the use of recursively generated intermediates that allow to utilize fast
matrix multiplication routines, can be found in Refs.~\cite{ccsd,nuclei6,gmatrix2,ccsdgrad}.
The analogous $m$-scheme-type expressions for the $\Theta_{i}^{a}(W_{N})$ and $\Theta_{ij}^{ab}(W_{N})$
contributions to the CCSD equations,
in terms of the $w_{stu}^{pqr}$ matrix elements defining $W_{N}$ and the
$t_{i}^{a}$ and $t_{ij}^{ab}$ amplitudes
can be found in Ref.~\cite{HaPa07}.
In using the CCSD equations presented in Refs.~\cite{ccsd,nuclei6,gmatrix2,ccsdgrad}, originally derived for two-body Hamiltonians, as expressions for $\Theta_{i}^{a}({\rm 2B})$ and $\Theta_{ij}^{ab}({\rm 2B})$ in the context of the calculations
including 3$N$ interactions, one only has to use  Eqs.~(\ref{fpq}) and~(\ref{vpqrs}) for the matrix elements $f_{q}^{p}$ and $v_{rs}^{pq}$ of the normal-ordered Hamiltonian, which contain the effective
$\half \sum_{i,j} h_{qij}^{pij}$ and $\qua \sum_{i} h_{rsi}^{pqi}$ contributions
due to the 3$N$ interactions. All of the remaining details are, however, the same. 
Following our earlier studies~\cite{RoBi12,HeBi13,BiLa13},
in performing the CCSD calculations for the closed-shell nuclei reported in this work, we 
use an angular-momentum-coupled formulation of CC theory discussed in Ref.~\cite{HaPa10},
which employs reduced matrix elements for all of the operators involved, allowing for a drastic reduction
in the number of matrix elements and cluster amplitudes entering the computations, and in a
substantial reduction in the number of CPU operations,
compared to a raw $m$-scheme description used in earlier nuclear CCSD 
work~\cite{nuclei1,gmatrix2,nuclei2,nuclei3,nuclei4,nuclei5,nuclei7,nuclei8,nuclei10,nuclei11},
enabling us to tackle medium-mass nuclei and larger numbers of
oscillator shells in the single-particle basis set.

Once the cluster amplitudes $t_{i}^{a}$ and $t_{ij}^{ab}$ are determined by solving the non-linear system represented by Eqs.~(\ref{ccsd:eqt1}) and~(\ref{ccsd:eqt2}), the CCSD correlation energy $\Delta E^{\rm (CCSD)}$, which is subsequently added to the
reference energy $\ERef$, Eq.~(\ref{ref-energy}), in order to obtain the total energy $E^{\rm (CCSD)}$,
as in Eq.~(\ref{ccsd-total-energy}),
is calculated using Eq.~(\ref{ccenergy}), where we replace
$\overline{H_{N}}$ by $\overline{H_{N}}^{\rm (CCSD)}$. We obtain
\beq
\Delta E^{\rm (CCSD)} = \Delta E_{\rm 2B}^{\rm (CCSD)} + \Delta E_{\rm 3B}^{\rm (CCSD)} ,
\label{ccsd-energy}
\eeq
where
\beq
\Delta E_{\rm 2B}^{\rm (CCSD)} = \langle \Phi | \overline{H_{N,{\rm 2B}}}^{\rm (CCSD)} | \Phi \rangle
\label{ccsd-energy-NO2B}
\eeq
and
\beq
\Delta E_{\rm 3B}^{\rm (CCSD)} =  \langle \Phi | \overline{W_{N}}^{\rm (CCSD)} | \Phi \rangle .
\label{ccsd-energy-WN}
\eeq
Again, in analogy to the standard two-body Hamiltonians, it is easy to show that
the NO2B contribution to the CCSD correlation energy, $\Delta E_{\rm 2B}^{\rm (CCSD)}$, can be
calculated using the expression
\begin{eqnarray}
\Delta E_{\rm 2B}^{\rm (CCSD)} & = & \langle \Phi | [ H_{N,{\rm 2B}} \, (T_{1} + T_{2} + \half T_{1}^{2}) ]_{C} | \Phi \rangle
\nonumber \\
& = & f_{a}^{i} t_{i}^{a} +  v_{ab}^{ij} ( \qua t_{ij}^{ab} + \half t_{i}^{a} t_{j}^{b}) ,
\label{energy-NO2B}
\end{eqnarray}
where $f_{a}^{i}$ and $v_{ab}^{ij}$ are determined using Eqs.~(\ref{fpq}) and~(\ref{vpqrs}).
For the $\Delta E_{\rm 3B}^{\rm (CCSD)}$ component of the CCSD correlation energy due to the residual
three-body interaction term $W_{N}$, we can write~\cite{HaPa07}
\begin{eqnarray}
\Delta E_{\rm 3B}^{\rm (CCSD)} & = & \langle \Phi | [W_{N} \, (T_{1} T_{2} + \six T_{1}^{3}) ]_{C} | \Phi \rangle
\nonumber \\
& = &  w_{abc}^{ijk} (\qua t_{i}^{a} t_{jk}^{bc} + \six t_{i}^{a} t_{j}^{b} t_{k}^{c} ) .
\label{energy-WN}
\end{eqnarray}
As in the case of Eq.~(\ref{cc-property}) and
other similar expressions shown in the rest of this section,
we have used the Einstein summation convention over the repeated upper and lower indices in the above
energy formulas.

We now move to the left-eigenstate or $\Lambda$CCSD equations, which one solves after the determination of the
$T_{1}$ and $T_{2}$ clusters and the CCSD energy, and which have to be solved prior to the
determination of the $\Lambda$CCSD(T) (or CR-CC(2,3)) energy correction $\delta E^{\rm (T)}$, since, as further elaborated on below, 
the $T_{1}$, $T_{2}$, $\Lambda_{1}$ and $\Lambda_{2}$
operators enter the $\delta E^{\rm (T)}$ expressions.
We examine the $\Lambda$CCSD equations
in full detail here, since the programmable form of these equations
for the case of 3$N$ interactions in the Hamiltonian
has never been considered before.

The left-eigenstate CCSD equations for the $\lambda_{a}^{i}$ and $\lambda_{ab}^{ij}$
amplitudes defining $\Lambda_{1}$ and $\Lambda_{2}$ are obtained by replacing the exact
$\Lambda$ and $\overline{H_{N}}$ operators in Eq.~(\ref{lambdacceqs-better}) by their truncated
CCSD counterparts, $\Lambda^{\rm (CCSD)}$ and $\overline{H_{N}}^{\rm (CCSD)}$, Eqs.~(\ref{lamccsd}) and~(\ref{hbarccsd}),
and by limiting the right-hand projections on the excited determinants
$|\Phi_{i_{1} \ldots i_{n}}^{a_{1} \ldots a_{n}} \rangle$ in Eq.~(\ref{lambdacceqs-better}) to the
singly and doubly excited determinants $|\Phi_{i}^{a}\rangle$ and $|\Phi_{ij}^{ab}\rangle$. This leads
to the following linear system for the $\Lambda_{1}$ and $\Lambda_{2}$ amplitudes (cf., e.g., 
Refs.~\cite{eomcc3,pp_rjb,gauss,ptcp2006,bartlett-musial2007,nucbook1,nuclei10,jspp-chemphys2012}):
\beq
\langle \Phi| (1+\Lambda_{1}+\Lambda_{2}) \, (\overline{H_{N}}^{\rm (CCSD)})_{\rm open} |\Phi_{i}^{a}\rangle = 0 ,
\label{ccsd:eqlambda1}
\eeq
\beq
\langle \Phi| (1+\Lambda_{1}+\Lambda_{2}) \, (\overline{H_{N}}^{\rm (CCSD)})_{\rm open} |\Phi_{ij}^{ab}\rangle = 0 .
\label{ccsd:eqlambda2}
\eeq
If we further split the similarity-transformed Hamiltonian of CCSD,
$\overline{H_{N}}^{\rm (CCSD)}$, into the NO2B and $W_{N}$ contributions
$\overline{H_{N,{\rm 2B}}}^{\rm (CCSD)}$ and $\overline{W_{N}}^{\rm (CCSD)}$, we can rewrite the $\Lambda$CCSD equations~(\ref{ccsd:eqlambda1}) and~(\ref{ccsd:eqlambda2}) for Hamiltonians including
three-body interactions as
\beq
\Xi_{a}^{i}({\rm 2B}) + \Xi_{a}^{i}(W_{N}) = 0 ,
\label{ccsd-lambda1}
\eeq
\beq
\Xi_{ab}^{ij}({\rm 2B}) + \Xi_{ab}^{ij}(W_{N}) = 0 ,
\label{ccsd-lambda2}
\eeq
where we define the corresponding NO2B and residual 3$N$ contributions as
\beq
\Xi_{a}^{i}({\rm 2B}) = \langle \Phi| (1+\Lambda_{1}+\Lambda_{2}) \, (\overline{H_{N,{\rm 2B}}}^{\rm (CCSD)})_{\rm open} |\Phi_{i}^{a}\rangle ,
\label{xi1-NO2B}
\eeq
\beq
\Xi_{a}^{i}(W_{N}) = \langle \Phi| (1+\Lambda_{1}+\Lambda_{2}) \, (\overline{W_{N}}^{\rm (CCSD)})_{\rm open} |\Phi_{i}^{a}\rangle ,
\label{xi1-WN}
\eeq
\beq
\Xi_{ab}^{ij}({\rm 2B}) = \langle \Phi| (1+\Lambda_{1}+\Lambda_{2}) \, (\overline{H_{N,{\rm 2B}}}^{\rm (CCSD)})_{\rm open} |\Phi_{ij}^{ab}\rangle ,
\label{xi2-NO2B}
\eeq
and
\beq
\Xi_{ab}^{ij}(W_{N}) = \langle \Phi| (1+\Lambda_{1}+\Lambda_{2}) \, (\overline{W_{N}}^{\rm (CCSD)})_{\rm open} |\Phi_{ij}^{ab}\rangle .
\label{xi2-WN}
\eeq
After identifying the non-vanishing terms in the above formulas and expressing them in terms of the
individual $n$-body components of the $\overline{H_{N,{\rm 2B}}}^{\rm (CCSD)}$ and
$\overline{W_{N}}^{\rm (CCSD)}$ operators, designated
in analogy to Eqs.~(\ref{HN-NO2B-many-body}) and~(\ref{WN-many-body}) by $\overline{H}_{n,{\rm 2B}}$ and
$\overline{W}_{n}$, we can write
\begin{eqnarray}
\Xi_{a}^{i}({\rm 2B}) & = & \langle \Phi|\{ [(1+\Lambda_{1}) \overline{H}_{1,{\rm 2B}}]_{C}
+ [(\Lambda_{1} + \Lambda_{2}) \overline{H}_{2,{\rm 2B}}]_{C}
\nonumber \\
&&
+ (\Lambda_{2} \overline{H}_{3,{\rm 2B}})_{C} \} |\Phi_{i}^{a}\rangle ,
\label{xi1-more-NO2B}
\end{eqnarray}
\begin{eqnarray}
\Xi_{ab}^{ij}({\rm 2B}) & = & \langle \Phi|\{ [(1+\Lambda_{1}+\Lambda_{2}) \overline{H}_{2,{\rm 2B}}]_{C}
+ (\Lambda_{2} \overline{H}_{1,{\rm 2B}})_{C}
\nonumber \\
&&
+ (\Lambda_{1} \overline{H}_{1,{\rm 2B}})_{DC}
+ (\Lambda_{2} \overline{H}_{3,{\rm 2B}})_{C} \} |\Phi_{ij}^{ab}\rangle ,
\label{xi2-more-NO2B}
\end{eqnarray}
\begin{eqnarray}
\Xi_{a}^{i}(W_{N}) & = & \langle \Phi| \{ [(1+\Lambda_{1}) \overline{W}_{1}]_{C}
+ [(\Lambda_{1} + \Lambda_{2}) \overline{W}_{2}]_{C}
\nonumber \\
&&
+ (\Lambda_{2} \overline{W}_{3})_{C} \} |\Phi_{i}^{a}\rangle ,
\label{xi1-more-WN}
\end{eqnarray}
and
\begin{eqnarray}
\Xi_{ab}^{ij}(W_{N}) & = & \langle \Phi|\{ [(1+\Lambda_{1}+\Lambda_{2}) \overline{W}_{2}]_{C}
+ (\Lambda_{2} \overline{W}_{1})_{C}
\nonumber \\
&&
+ (\Lambda_{1} \overline{W}_{1})_{DC}
+ [(\Lambda_{1}+\Lambda_{2}) \overline{W}_{3}]_{C}
\nonumber \\
&&
+ (\Lambda_{2} \overline{W}_{4})_{C} \}
|\Phi_{ij}^{ab}\rangle ,
\label{xi2-more-WN}
\end{eqnarray}
where $C$ continues to represent the connected operator product and
$DC$ stands for the disconnected product expression. The detailed $m$-scheme-style
formulas for the $\Xi^i_a({\rm 2B})$, $\Xi^{ij}_{ab}({\rm 2B})$, $\Xi^i_a(W_{N})$, and $\Xi^{ij}_{ab}(W_{N})$
contributions to the $\Lambda$CCSD system represented by Eqs. (\ref{ccsd-lambda1}) and (\ref{ccsd-lambda2}),
in terms of the individual matrix elements
$\helem{p_{1} \ldots p_{n}}{q_{1} \ldots q_{n}}$ and $\welem{p_{1} \ldots p_{n}}{q_{1} \ldots q_{n}}$
that define the $n$-body components of $\overline{H_{N,{\rm 2B}}}^{\rm (CCSD)}$ and
$\overline{W_{N}}^{\rm (CCSD)}$ 
are given by
\begin{eqnarray}
\Xi_a^i({\rm 2B}) & = &
\helem{i}{a}
+ \lambda^i_c \, \helem{c}{a}
- \lambda^k_a \, \helem{i}{k}
\nonumber \\
&&
+ \lambda^k_c \, \helem{ci}{ka}
+ \tfrac{1}{2} \lambda^{ik}_{cd} \, \helem{cd}{ak}
\nonumber \\
&&
- \tfrac{1}{2} \lambda^{kl}_{ac} \, \helem{ic}{kl}
+ \tfrac{1}{4} \lambda^{kl}_{cd} \, \helem{cdi}{kla} ,
\label{xi-ia-2B} \\[10pt]
\Xi_{ab}^{ij}({\rm 2B}) & = &
\helem{ij}{ab}
+ {\mathscr A}_{ab} {\mathscr A}^{ij} \lambda^j_b \, \helem{i}{a}
\nonumber \\
&&
+ {\mathscr A}^{ij} \lambda^i_c \, \helem{cj}{ab}
- {\mathscr A}_{ab} \lambda^k_a \, \helem{ij}{kb}
\nonumber \\
&&
+ {\mathscr A}_{ab} \lambda^{ij}_{ac} \, \helem{c}{b}
- {\mathscr A}^{ij} \lambda^{ik}_{ab} \, \helem{j}{k}
\nonumber \\
&&
+ {\mathscr A}_{ab} {\mathscr A}^{ij} \lambda^{ik}_{ac} \, \helem{cj}{kb}
+ \tfrac{1}{2} \lambda^{ij}_{cd} \, \helem{cd}{ab}
\nonumber \\
&&
+ \tfrac{1}{2} \lambda^{kl}_{ab} \, \helem{ij}{kl}
+ \tfrac{1}{2} {\mathscr A}_{ab} \lambda^{kl}_{ca} \, \helem{ijc}{kbl}
\nonumber \\
&&
+ \tfrac{1}{2} {\mathscr A}^{ij} \lambda^{ki}_{cd} \, \helem{cjd}{abk} ,
\label{xi-ijab-2B} \\[10pt]
\Xi_a^i(W_N) & = &
\welem{i}{a}
+ \lambda^i_c \, \welem{c}{a}
- \lambda^k_a \, \welem{i}{k}
+ \lambda^k_c \, \welem{ci}{ka}
\nonumber \\
&&
+ \tfrac{1}{2} \lambda^{ik}_{cd} \, \welem{cd}{ak}
- \tfrac{1}{2} \lambda^{kl}_{ac} \, \welem{ic}{kl}
+ \tfrac{1}{4} \lambda^{kl}_{cd} \, \welem{cdi}{kla} ,
\label{xi-ia-WN}
\end{eqnarray}
and
\begin{eqnarray}
\Xi_{ab}^{ij}(W_N) & = &
\welem{ij}{ab}
+ {\mathscr A}_{ab} {\mathscr A}^{ij} \lambda^j_b \, \welem{i}{a}
+ {\mathscr A}^{ij} \lambda^i_c \, \welem{cj}{ab}
\nonumber \\
&&
- {\mathscr A}_{ab} \lambda^k_a \, \welem{ij}{kb}
+ \lambda^{k}_{c} \, \welem{ijc}{abk}
+ {\mathscr A}_{ab} \lambda^{ij}_{ac} \, \welem{c}{b}
\nonumber \\
&&
- {\mathscr A}^{ij} \lambda^{ik}_{ab} \, \welem{j}{k}
+ {\mathscr A}_{ab} {\mathscr A}^{ij} \lambda^{ik}_{ac} \, \welem{cj}{kb}
\nonumber \\
&&
+ \tfrac{1}{2} \lambda^{ij}_{cd} \, \welem{cd}{ab}
+ \tfrac{1}{2} \lambda^{kl}_{ab} \, \welem{ij}{kl}
+ \tfrac{1}{2} {\mathscr A}_{ab} \lambda^{kl}_{ca} \, \welem{ijc}{kbl}
\nonumber \\
&&
+ \tfrac{1}{2} {\mathscr A}^{ij} \lambda^{ki}_{cd} \, \welem{cjd}{abk}
+ \tfrac{1}{4} \lambda^{kl}_{cd} \welem{ijcd}{abkl} ,
\label{xi-ijab-WN}
\end{eqnarray}
respectively, where
\beq
{\mathscr A}_{pq} \equiv {\mathscr A}^{pq} = 1 - (pq),
\label{antisymm-pq}
\eeq
with $(pq)$ representing a transposition of $p$ and $q$, are the usual index antisymmetrizers.

As one can see, the $\Lambda$CCSD equations for three-body Hamiltonians, although more complicated
than for the case of pairwise interactions, where one would not consider 
Eqs.~(\ref{xi-ia-WN}) and~(\ref{xi-ijab-WN}), have a relatively simple algebraic
structure. In particular, the highest-rank many-body components of the
$\overline{H_{N,{\rm 2B}}}^{\rm (CCSD)}$ and $\overline{W_{N}}^{\rm (CCSD)}$
operators that enter these equations are given by selected types of
three-body $\overline{H}_{3,{\rm 2B}}$ terms and selected types of 
four-body $\overline{W}_{4}$ terms. Although, according to the remarks below
Eqs.~(\ref{HN-NO2B-many-body})--(\ref{WN-n}), the $\overline{H_{N,{\rm 2B}}}^{\rm (CCSD)}$
and $\overline{W_{N}}^{\rm (CCSD)}$ operators contain various higher--than--four-body terms,
the right-hand projections on the singly and doubly excited determinants in
Eqs.~(\ref{ccsd:eqlambda1}) and~(\ref{ccsd:eqlambda2}) 
or~(\ref{xi1-NO2B})--(\ref{xi2-WN}) eliminate such complicated expressions.
This greatly simplifies the computer implementation effort.
Again, in performing the left-eigenstate CCSD
calculations for the closed-shell nuclei reported in this work, following the recipe presented in Ref.~\cite{HaPa10}, 
we convert the $m$-scheme expressions
for the $\Xi_{a}^{i}({\rm 2B})$, $\Xi_{ab}^{ij}({\rm 2B})$, $\Xi_{a}^{i}(W_{N})$, and $\Xi_{ab}^{ij}(W_{N})$
contributions into their angular-momentum-coupled representation.
 The key quantities for setting up the underlying
Eqs.~(\ref{xi-ia-2B})--(\ref{xi-ijab-WN}) are the matrix elements
$\helem{q_{1} \ldots q_{n}}{p_{1} \ldots p_{n}}$ and $\welem{q_{1} \ldots q_{n}}{p_{1} \ldots p_{n}}$
of the similarity-transformed $\overline{H_{N,{\rm 2B}}}^{\rm (CCSD)}$ and
$\overline{W_{N}}^{\rm (CCSD)}$ operators. Before discussing the sources of information about the
matrix elements of $\overline{H_{N,{\rm 2B}}}^{\rm (CCSD)}$ and $\overline{W_{N}}^{\rm (CCSD)}$
that enter Eqs.~(\ref{xi-ia-2B})--(\ref{xi-ijab-WN}), let us comment
on the physical and mathematical content of these equations, including important additional
simplifications in the NO2B contributions $\Xi_{a}^{i}({\rm 2B})$ and $\Xi_{ab}^{ij}({\rm 2B})$
that reduce the usage of higher--than--two-body objects in the equations for
the $\lambda_{a}^{i}$ and $\lambda_{ab}^{ij}$ amplitudes even further.

First, we note that the NO2B and residual 3$N$ components of the $\Lambda$CCSD equations
projected on the singly excited $|\Phi_{i}^{a}\rangle$ determinants,
$\Xi_{a}^{i}({\rm 2B})$ and $\Xi_{a}^{i}(W_{N})$, have the identical
general form, i.e., they only differ by the details of the
Hamiltonian matrix elements that enter them, but not by their
overall algebraic structure (cf. Eqs.~(\ref{xi1-more-NO2B}) or~(\ref{xi-ia-2B}) 
and~(\ref{xi1-more-WN}) or~(\ref{xi-ia-WN})). However, in the NO2B case,
the contribution
\beq
\langle \Phi| (\Lambda_{2} \overline{H}_{3,{\rm 2B}})_{C} |\Phi_{i}^{a}\rangle = \tfrac{1}{4} \lambda^{kl}_{cd} \, \helem{cdi}{kla} ,
\label{3body-singles-2B}
\eeq
which contains selected three-body components of $\overline{H_{N,{\rm 2B}}}^{\rm (CCSD)}$ and
which enters Eqs.~(\ref{xi1-more-NO2B}) and~(\ref{xi-ia-2B}) for $\Xi_{a}^{i}({\rm 2B})$,
can be refactorized and rewritten in terms of simpler one- and two-body objects,
eliminating the need for the explicit use of the three-body $\overline{H}_{3,{\rm 2B}}$ terms altogether. Indeed, following
the quantum-chemistry literature where interactions in the Hamiltonian are always two-body, we can
replace Eq.~(\ref{3body-singles-2B}) by
(cf., e.g., Ref.~\cite{ccsdgrad})
\beq
\tfrac{1}{4} \lambda^{kl}_{cd} \, \helem{cdi}{kla}
=
- \helem{ie}{ad} \, \chi^d_e 
- \helem{im}{an} \, \chi^n_m ,
\label{3body-singles-alt}
\eeq
where the additional one-body intermediates $\chi^d_e$ and $\chi^n_m$ are defined as
\beq
\chi^d_e = -\tfrac{1}{2} t^{df}_{mn} \lambda^{mn}_{ef}
\label{chi-de}
\eeq
and
\beq
\chi^l_m = \tfrac{1}{2} t^{ef}_{mn} \lambda^{ln}_{ef} ,
\label{chi-lm}
\eeq
respectively.
In other words, all we need to know to construct the NO2B contribution $\Xi_{a}^{i}({\rm 2B})$ to
the $\Lambda$CCSD equations are the matrix elements $\helem{p}{q}$ and $\helem{pq}{rs}$ of the
similarity-transformed Hamiltonian $\overline{H_{N,{\rm 2B}}}^{\rm (CCSD)}$,
which appear in Eqs.~(\ref{xi-ia-2B}) and~(\ref{3body-singles-alt}),
and the 
cluster amplitudes $t_{i}^{a}$ and $t_{ij}^{ab}$,
plus two auxiliary one-body intermediates, obtained by contracting the
$t_{ij}^{ab}$ and $\lambda_{ab}^{ij}$ amplitudes,
defined by Eqs.~(\ref{chi-de}) and~(\ref{chi-lm}).
The relevant, computationally efficient, expressions for the
one- and two-body matrix elements $\helem{p}{q}$ and $\helem{pq}{rs}$ can be found
in several sources, for example in Refs.~\cite{nuclei6,crccl_ijqc2,creomopen},
remembering to rely on Eqs.~(\ref{fpq}) and~(\ref{vpqrs})
in the determination of $f_{q}^{p}$ and $v_{rs}^{pq}$.
Unfortunately, we cannot provide any additional simplifications in the case of the $W_{N}$
analog of Eq.~(\ref{3body-singles-2B}), entering Eqs.~(\ref{xi1-more-WN}) and~(\ref{xi-ia-WN}), 
\beq
\langle \Phi| (\Lambda_{2} \overline{W}_{3})_{C} |\Phi_{i}^{a}\rangle = \tfrac{1}{4} \lambda^{kl}_{cd} \, \welem{cdi}{kla} ,
\label{3body-singles-WN}
\eeq
where we have to rely on the intrinsically three-body matrix elements of $W_{N}$
that do not factorize into simpler, lower-rank objects.
In this case, in order to construct the residual 3$N$ contribution $\Xi_{a}^{i}(W_{N})$ to
the $\Lambda$CCSD equations projected on $|\Phi_{i}^{a}\rangle$, given by
Eq.~(\ref{xi-ia-WN}), we must utilize the explicit formulas for the one-, two-, and three-body
matrix elements of the similarity-transformed $\overline{W_{N}}^{\rm (CCSD)}$ operator
in terms of the appropriate matrix elements $w_{stu}^{pqr}$ of $W_{N}$ and
the CCSD amplitudes $t_{i}^{a}$ and $t_{ij}^{ab}$ that are listed in 
Tables~\ref{tab:EffectiveHamiltonian1B2B} and~\ref{tab:EffectiveHamiltonian3B}.

Similar, albeit not identical, remarks apply to the $\Lambda$CCSD equations projected on
the doubly excited determinants $|\Phi_{ij}^{ab}\rangle$. Once again, we can refactorize
the NO2B contribution
\begin{eqnarray}
\langle \Phi| (\Lambda_{2} \overline{H}_{3,{\rm 2B}})_{C} |\Phi_{ij}^{ab}\rangle
& = & \tfrac{1}{2} {\mathscr A}_{ab} \lambda^{kl}_{ca} \, \helem{ijc}{kbl}
\nonumber \\
&&
+ \tfrac{1}{2} {\mathscr A}^{ij} \lambda^{ki}_{cd} \, \helem{cjd}{abk} ,
\label{3body-doubles-2B}
\end{eqnarray}
entering Eqs.~(\ref{xi2-more-NO2B}) and~(\ref{xi-ijab-2B}),
which contains selected three-body components of $\overline{H_{N,{\rm 2B}}}^{\rm (CCSD)}$,
by rewriting it in terms of simpler one- and two-body objects as
\begin{eqnarray}
\tfrac{1}{2} {\mathscr A}_{ab} \lambda^{kl}_{ca} \, \helem{ijc}{kbl}
& + & \tfrac{1}{2} {\mathscr A}^{ij} \lambda^{ki}_{cd} \, \helem{cjd}{abk} 
\nonumber \\
& = &
{\mathscr A}_{ab} \helem{ij}{ad} \, \chi^d_b
- {\mathscr A}^{ij} \helem{im}{ab} \, \chi^j_m
\nonumber \\
& = &
{\mathscr A}_{ab} v^{ij}_{ad} \, \chi^d_b
- {\mathscr A}^{ij} v^{im}_{ab} \, \chi^j_m ,
\label{3body-doubles-alt}
\end{eqnarray}
using the identity $\helem{kl}{cd} = v^{kl}_{cd}$ and
where $\chi^d_b$ and $\chi^j_m$ are again given by Eqs.~(\ref{chi-de}) and~(\ref{chi-lm}),
but we cannot do anything similar for the case of the analogous expression
\beq
\langle \Phi| (\Lambda_{2} \overline{W}_{3})_{C} |\Phi_{ij}^{ab}\rangle
= \tfrac{1}{2} {\mathscr A}_{ab} \lambda^{kl}_{ca} \, \welem{ijc}{kbl}
+ \tfrac{1}{2} {\mathscr A}^{ij} \lambda^{ki}_{cd} \, \welem{cjd}{abk}
\eeq
 that appears in  Eqs.~(\ref{xi2-more-WN}) and~(\ref{xi-ijab-WN}),
where we have to rely on the three-body matrix elements of $W_{N}$.
As a result, in analogy to the previously examined $\Xi_{a}^{i}({\rm 2B})$ term,
all we need to know to construct the NO2B contribution $\Xi_{ab}^{ij}({\rm 2B})$ to
the $\Lambda$CCSD equations 
are the matrix elements $\helem{p}{q}$ and $\helem{pq}{rs}$
of $\overline{H_{N,{\rm 2B}}}^{\rm (CCSD)}$, 
plus two auxiliary one-body intermediates
defined by Eqs.~(\ref{chi-de}) and~(\ref{chi-lm}),
but one needs additional expressions for the various matrix elements of
$\overline{W_{N}}^{\rm (CCSD)}$
to construct $\Xi_{ab}^{ij}(W_{N})$, Eq.~(\ref{xi-ijab-WN}). In fact,
the situation with the residual $W_{N}$ contributions to the
$\Lambda$CCSD equations projected on $|\Phi_{ij}^{ab}\rangle$
is further complicated by the observation that along with the various terms that are analogous
to the NO2B case, we also end up with the additional
\beq
\langle \Phi| (\Lambda_{1} \overline{W}_{3})_{C} |\Phi_{ij}^{ab}\rangle = \lambda^{k}_{c} \, \welem{ijc}{abk}
\label{extra3}
\eeq
and
\beq
\langle \Phi| (\Lambda_{2} \overline{W}_{4})_{C} |\Phi_{ij}^{ab}\rangle = \tfrac{1}{4} \lambda^{kl}_{cd} \welem{ijcd}{abkl}
\label{extra4}
\eeq
contributions to $\Xi_{ab}^{ij}(W_{N})$, which contain selected three- and four-body
components of $\overline{W_{N}}^{\rm (CCSD)}$ and which
do not have their NO2B equivalents in $\Xi_{ab}^{ij}({\rm 2B})$
(cf. Eqs.~(\ref{xi2-more-NO2B}) or~(\ref{xi-ijab-2B}) 
and~(\ref{xi2-more-WN}) or~(\ref{xi-ijab-WN})),
since one cannot form such terms from two-body Hamiltonians.
%
The former term, Eq.~(\ref{extra3}), cannot be further simplified, but the latter contribution can be expressed in a computationally efficient, factorized form utilizing the previously defined intermediates given by Eqs.~(\ref{chi-de}) and~(\ref{chi-lm}), obtaining
\begin{eqnarray}
\langle \Phi| (\Lambda_{2} \overline{W}_{4})_{C} |\Phi_{ij}^{ab}\rangle
&=&
- (\welem{ijc}{abd} - \welem{ijm}{abd} t^c_m ) \, \chi^d_c \nonumber \\
&&
- (\welem{ijl}{abk} + \welem{ijl}{abd} t^d_k ) \, \chi^k_l .
\end{eqnarray}

The complete set of expressions for the one-, two-, three-, and
four-body matrix elements of $\overline{W_{N}}^{\rm (CCSD)}$, in terms of the
pertinent $w_{stu}^{pqr}$ matrix elements of $W_{N}$ and the
CCSD amplitudes $t_{i}^{a}$ and $t_{ij}^{ab}$ is given in 
Tables~\ref{tab:EffectiveHamiltonian1B2B} and~\ref{tab:EffectiveHamiltonian3B}.

\begin{table}[h]
\begin{eqnarray}
\welem{i}{a} &=& 
\tfrac{1}{4} 
\mWME{ikl}{acd} 
t^{cd}_{kl}
+
\tfrac{1}{2}
\mWME{ikl}{acd} 
t^c_k 
t^d_l
\nonumber \\
\welem{a}{b} &=&
\tfrac{1}{4}
\mWME{akl}{bcd} 
t^{cd}_{kl} 
-
\tfrac{1}{4} 
\mWME{klm}{bcd} 
t^{cd}_{kl} 
t^a_m 
+
\tfrac{1}{2} 
\mWME{klm}{bcd} 
t^c_k 
t^{ad}_{lm}
+
\tfrac{1}{2}
\mWME{akl}{bcd} 
t^c_k 
t^d_l
\nonumber \\
&&
-
\tfrac{1}{2}
\mWME{klm}{bcd} 
t^c_k 
t^d_l 
t^a_m 
\nonumber \\
\welem{i}{j} &=& 
\tfrac{1}{4} 
\mWME{ikl}{cdj} 
t^{cd}_{kl}
\nonumber
+
\tfrac{1}{4}
\mWME{ikl}{cde} 
t^{cd}_{kl} 
t^e_j
-
\tfrac{1}{2}
\mWME{ikl}{cde} 
t^{cd}_{kj} 
t^e_l
+
\tfrac{1}{2}
\mWME{ikl}{cdj} 
t^c_k 
t^d_l
\nonumber \\
&&
+
\tfrac{1}{2}
\mWME{ikl}{cde} 
t^c_k 
t^d_l 
t^e_j
\nonumber \\
\welem{ij}{ab} &=& 
\mWME{ijk}{abc} 
t^c_k
\nonumber \\
\welem{ai}{bc} &=& 
\tfrac{1}{2}
\mWME{ikl}{bcd} 
t^{ad}_{kl} 
+
\mWME{ail}{bcd} 
t^d_l 
+
\mWME{ikl}{bcd} 
t^a_k 
t^d_l 
\nonumber \\
\welem{ik}{ja} &=& 
-
\tfrac{1}{2}
\mWME{ikl}{acd} 
t^{cd}_{jl} 
+
\mWME{ikl}{jac} 
t^c_l 
-
\mWME{ikl}{acd} 
t^c_j 
t^d_l 
\nonumber \\
\welem{ab}{cd} &=&
-
\tfrac{1}{2} 
{\mathscr A}^{ab} 
\mWME{akl}{cde} 
t^{be}_{kl} 
+
\mWME{abk}{cde} 
t^e_k \
+
\tfrac{1}{2}
\mWME{klm}{cde} 
t^{ab}_{lm} 
t^e_k 
+
\tfrac{1}{2} 
{\mathscr A}^{ab}
\mWME{klm}{cde} 
t^a_k 
t^{be}_{lm} 
\nonumber \\
&&
-
\mWME{bkl}{cde} 
t^a_k 
t^e_l 
+
\mWME{klm}{cde} 
t^a_l 
t^b_m 
t^e_k 
\nonumber \\
\welem{ij}{kl} &=&
-
\tfrac{1}{2} 
{\mathscr A}_{kl}
\mWME{ijm}{kcd} 
t^{cd}_{ml} 
+
\mWME{ijm}{klc} 
t^c_m 
+
\tfrac{1}{2}
\mWME{ijm}{cde} 
t^{de}_{kl} 
t^c_m 
+
\tfrac{1}{2} 
{\mathscr A}_{kl}
\mWME{ijm}{cde} 
t^{cd}_{lm} 
t^e_k 
\nonumber \\
&&
+
\mWME{ijm}{cdl} 
t^c_k 
t^d_m 
+
\mWME{ijm}{cde} 
t^d_k 
t^e_l 
t^c_m 
\nonumber \\
\welem{aj}{ib} &=& 
-
\tfrac{1}{2}
\mWME{ajk}{bcd} 
t^{cd}_{ik} 
+
\tfrac{1}{2}
\mWME{jkl}{bci} 
t^{ac}_{kl} 
+
\mWME{ajk}{bci} 
t^c_k 
-
\tfrac{1}{2}
\mWME{jkl}{bcd} 
t^{cd}_{il} 
t^a_k 
-
\tfrac{1}{2}
\mWME{jkl}{bcd} 
t^{ad}_{kl} 
t^c_i
\nonumber \\
&&
+
\mWME{jkl}{bcd} 
t^{ac}_{ik} 
t^d_l 
+
\mWME{ajk}{bcd} 
t^c_k 
t^d_i 
-
\mWME{jkl}{bci} 
t^c_k 
t^a_l 
-
\mWME{jkl}{bcd} 
t^a_l 
t^c_k 
t^d_i 
\nonumber \\
\welem{ab}{ci} &=& 
-
\mWME{abk}{cdi} 
t^d_k 
+
\tfrac{1}{2} 
{\mathscr A}^{ab}
\mWME{akl}{cde} 
t^{de}_{ik} 
t^b_l 
+
\tfrac{1}{2} 
{\mathscr A}^{ab}
\mWME{akl}{cde} 
t^{bd}_{kl} 
t^e_i 
+
{\mathscr A}^{ab}
\mWME{akl}{cde} 
t^{be}_{il} 
t^d_k 
\nonumber \\
&&
-
\tfrac{1}{2}
\mWME{klm}{cdi} 
t^{ab}_{lm} 
t^d_k 
-
\tfrac{1}{2} 
{\mathscr A}^{ab}
\mWME{klm}{cdi} 
t^{bd}_{kl} 
t^a_m 
-
\mWME{abk}{cde} 
t^d_k 
t^e_i
+
{\mathscr A}^{ab}
\mWME{bkl}{cdi} 
t^d_k 
t^a_l 
\nonumber \\
&&
+
\tfrac{1}{2}
\mWME{klm}{cde} 
t^{ab}_{im} 
t^d_k 
t^e_l 
-
{\mathscr A}^{ab}
\mWME{klm}{cde} 
t^{be}_{il} 
t^d_k 
t^a_m
+
\tfrac{1}{2} 
\mWME{klm}{cde} 
t^{de}_{il} 
t^b_k 
t^a_m 
\nonumber \\
&&
-
\tfrac{1}{2} 
{\mathscr A}^{ab}
\mWME{klm}{cde} 
t^{ae}_{lm} 
t^b_k 
t^d_i 
-
\tfrac{1}{2}
\mWME{klm}{cde} 
t^{ab}_{lm} 
t^d_k 
t^e_i 
-
{\mathscr A}^{ab}
\mWME{akl}{cde} 
t^d_k 
t^b_l 
t^e_i 
\nonumber \\
&&
+
\mWME{klm}{cdi} 
t^a_m 
t^b_l 
t^d_k
+
\mWME{klm}{cde} 
t^a_m 
t^b_l 
t^d_k 
t^e_i 
-
\tfrac{1}{4}
\mWME{klm}{cde} 
t^{ab}_{ki} 
t^{de}_{lm} 
+
\tfrac{1}{4}
\mWME{klm}{cde} 
t^{ab}_{kl} 
t^{de}_{im} 
\nonumber \\
&&
-
\tfrac{1}{2} 
{\mathscr A}^{ab}
\mWME{klm}{cde} 
t^{ad}_{kl} 
t^{eb}_{mi}
\nonumber \\
\welem{ia}{jk} &=&
\mWME{ial}{jkc} 
t^c_l 
+
\tfrac{1}{2} 
{\mathscr A}_{jk}
\mWME{ilm}{jcd} 
t^{ac}_{lm} 
t^d_k 
+
\tfrac{1}{2} 
{\mathscr A}_{jk}
\mWME{ilm}{jcd} 
t^{cd}_{kl} 
t^a_m 
+
{\mathscr A}_{jk}
\mWME{ilm}{jcd} 
t^{ad}_{km} 
t^c_l
\nonumber \\
&&
-
\tfrac{1}{2}
\mWME{ail}{cde} 
t^{de}_{jk} 
t^c_l
+
\tfrac{1}{2} 
{\mathscr A}_{jk}
\mWME{ail}{cde} 
t^{cd}_{lk} 
t^e_j 
-
\mWME{ilm}{jck} 
t^c_l 
t^a_m 
-
{\mathscr A}_{jk}
\mWME{ail}{cdk} 
t^c_l 
t^d_j
\nonumber \\
&&
+
\tfrac{1}{2} 
\mWME{ilm}{cde} 
t^{ae}_{kj} 
t^c_l 
t^d_m 
+
{\mathscr A}_{jk}
\mWME{ilm}{cde}
t^{ad}_{km} 
t^c_l 
t^e_j 
-
\mWME{ilm}{cde} 
t^{ad}_{lm} 
t^c_k 
t^e_j 
\nonumber \\
&&
-
\tfrac{1}{2} 
{\mathscr A}_{jk}
\mWME{ilm}{cde} 
t^{de}_{mj} 
t^c_k 
t^a_l 
+
{\mathscr A}_{jk}
\mWME{ilm}{cde} 
t^{ae}_{jm} 
t^c_l 
t^d_k 
-
{\mathscr A}_{jk}
\mWME{ilm}{jcd} 
t^a_m 
t^c_l 
t^d_k 
\nonumber \\
&&
+
\mWME{ail}{cde} 
t^c_l 
t^d_k 
t^e_j 
-
\mWME{ilm}{cde} 
t^a_m 
t^c_l 
t^d_k 
t^e_j 
+
\tfrac{1}{4}
\mWME{ilm}{cde} 
t^{ca}_{jk} 
t^{de}_{lm} 
-
\tfrac{1}{4}
\mWME{ilm}{cde} 
t^{cd}_{jk} 
t^{ae}_{lm} 
\nonumber \\
&&
+
\tfrac{1}{2} 
{\mathscr A}_{jk}
\mWME{ilm}{cde} 
t^{cd}_{jl} 
t^{ea}_{mk} 
\nonumber
\end{eqnarray}

\caption{Explicit expressions for the one- and two-body matrix elements of the similarity-transformed form of the
the residual three-body interaction term $W_{N}$, designated by $\overline{W_{N}}^{\rm (CCSD)}$ and defined by Eq.~(\ref{sim-WN-ccsd}),
which are needed to construct the $\Xi_{a}^{i}(W_{N})$ and $\Xi_{ab}^{ij}(W_{N})$ contributions to the $\Lambda$CCSD equations,
Eqs.~(\ref{xi-ia-WN}) and~(\ref{xi-ijab-WN}), respectively.}
\label{tab:EffectiveHamiltonian1B2B}
\end{table}

\begin{table}[h]

\begin{eqnarray}
\welem{ija}{kbl} &=&
\mWME{ija}{kbl} \
+
{\mathscr A}_{kl}
\mWME{ijm}{kbc} 
t^{ac}_{lm} 
+
\tfrac{1}{2} 
\mWME{ija}{cbd} 
t^{cd}_{kl} 
-
{\mathscr A}_{kl}
\mWME{ija}{bcl} 
t^c_k
+
\mWME{ijm}{bkl} 
t^a_m
\nonumber \\
&&
-
\mWME{ijm}{bcd} 
t^{ad}_{kl} 
t^c_m
-
{\mathscr A}_{kl}
\mWME{ijm}{bcd} 
t^{ad}_{lm} 
t^c_k
+
\tfrac{1}{2} 
\mWME{ijm}{bcd} 
t^{cd}_{kl} 
t^a_m 
\nonumber \\
&&
+
{\mathscr A}_{kl}
\mWME{ijm}{bcl} 
t^c_k 
t^a_m
\nonumber 
+
\mWME{ija}{cbd} 
t^c_k 
t^d_l 
+
\mWME{ijm}{bcd} 
t^c_k 
t^d_l 
t^a_m
\nonumber \\
\welem{ajb}{cdi} &=& 
\mWME{ajb}{cdi} 
+
{\mathscr A}^{ab} 
\mWME{ajk}{cde} 
t^{be}_{ik}
+
\tfrac{1}{2} 
\mWME{kjl}{cdi} 
t^{ab}_{kl} 
-
{\mathscr A}^{ab} 
\mWME{bkj}{cdi} 
t^a_k 
\nonumber \\
&&
-
\mWME{abj}{cde} 
t^e_i
+
\mWME{jkl}{cde} 
t^{ab}_{il} 
t^e_k
+
{\mathscr A}^{ab} 
\mWME{jkl}{cde} 
t^{be}_{il} 
t^a_k
-
\tfrac{1}{2} 
\mWME{jkl}{cde} 
t^{ab}_{kl} 
t^e_i
\nonumber \\
&&
-
{\mathscr A}^{ab} 
\mWME{ajk}{cde} 
t^b_k 
t^e_i
+
\mWME{kjl}{cdi} 
t^a_k 
t^b_l 
-
\mWME{jkl}{cde} 
t^a_k 
t^b_l 
t^e_i
\nonumber \\
\welem{abk}{ijc} &=& 
\mWME{abk}{ijc} 
+
\tfrac{1}{2} 
\mWME{abk}{cde} 
t^{de}_{ij} 
+
{\mathscr A}^{ab} {\mathscr A}_{ij}
\mWME{bkl}{jcd} 
t^{ad}_{il} 
+
\tfrac{1}{2} 
\mWME{klm}{ijc} 
t^{ab}_{lm} 
\nonumber \\
&&
-
\tfrac{1}{2} 
{\mathscr A}^{ab}
\mWME{klm}{cde} 
t^{ad}_{ij} 
t^{be}_{lm} 
+
\tfrac{1}{2} 
{\mathscr A}^{ab} {\mathscr A}_{ij}
\mWME{klm}{cde} 
t^{ad}_{il} 
t^{be}_{jm} 
+
\tfrac{1}{4} 
\mWME{klm}{cde} 
t^{ab}_{lm} 
t^{de}_{ij} 
\nonumber \\
&&
-
\tfrac{1}{2} 
{\mathscr A}_{ij}
\mWME{klm}{cde} 
t^{ab}_{il} 
t^{de}_{jm} 
+
{\mathscr A}_{ij}
\mWME{abk}{cdj} 
t^{d}_{i} 
-
{\mathscr A}^{ab}
\mWME{klb}{ijc} 
t^{a}_{l} 
-
\tfrac{1}{2} 
{\mathscr A}^{ab}
\mWME{klb}{cde} 
t^{de}_{ij} 
t^{a}_{l} 
\nonumber \\
&&
-
{\mathscr A}^{ab} {\mathscr A}_{ij}
\mWME{akl}{cde} 
t^{be}_{jl} 
t^{d}_{i} 
+
\tfrac{1}{2} 
{\mathscr A}_{ij}
\mWME{klm}{cdj} 
t^{ab}_{lm} 
t^{d}_{i} 
+
{\mathscr A}^{ab} {\mathscr A}_{ij}
\mWME{klm}{cdi} 
t^{bd}_{jm} 
t^{a}_{l} 
\nonumber \\
&&
-
{\mathscr A}^{ab}
\mWME{akl}{cde} 
t^{be}_{ij} 
t^{d}_{l} 
+
{\mathscr A}_{ij}
\mWME{klm}{cdi} 
t^{ab}_{jm} 
t^{d}_{l} 
+
\mWME{abk}{cde}
t^{d}_{i}
t^{e}_{j}
-
{\mathscr A}^{ab} {\mathscr A}_{ij}
\mWME{bkl}{cdj}
t^{a}_{l}
t^{d}_{i}
\nonumber \\
&&
+
\mWME{klm}{ijc}
t^{a}_{l}
t^{b}_{m}
+
\tfrac{1}{2}
\mWME{klm}{cde}
t^{de}_{ij}
t^{a}_{l}
t^{b}_{m}
-
{\mathscr A}^{ab} {\mathscr A}_{ij}
\mWME{klm}{cde}
t^{be}_{jm}
t^{a}_{l}
t^{d}_{i}
+
\tfrac{1}{2}
\mWME{klm}{cde}
t^{ab}_{lm}
t^{d}_{i}
t^{e}_{j}
\nonumber \\
&&
-
{\mathscr A}^{ab}
\mWME{klm}{cde}
t^{eb}_{ij}
t^{a}_{m}
t^{d}_{l}
-
{\mathscr A}_{ij}
\mWME{klm}{cde}
t^{ab}_{mj}
t^{d}_{l}
t^{e}_{i}
-
{\mathscr A}^{ab}
\mWME{bkl}{cde}
t^{a}_{l}
t^{d}_{i}
t^{e}_{j}
\nonumber \\
&&
+
{\mathscr A}_{ij}
\mWME{klm}{cdj}
t^{a}_{l}
t^{b}_{m}
t^{d}_{i}
+
\mWME{klm}{cde}
t^{a}_{l}
t^{b}_{m}
t^{d}_{i}
t^{e}_{j}
\nonumber \\
\welem{ijc}{abk} &=&
\mWME{ijc}{abk}
+
\mWME{ijc}{abd}
t^d_k
-
\mWME{ijl}{abk}
t^c_l
-
\mWME{ijl}{abd}
t^c_l
t^d_k
+
\mWME{ijl}{abd}
t^{cd}_{kl}
\nonumber \\
\welem{ijcd}{abkl} &=&
{\mathscr A}^{cd}
\mWME{ijc}{abe} 
t^{ed}_{kl}
-
{\mathscr A}_{kl}
\mWME{ijm}{abk} 
t^{cd}_{ml} 
-
{\mathscr A}^{cd}
\mWME{ijm}{abe} 
t^{ed}_{kl} 
t^c_m
-
{\mathscr A}_{kl}
\mWME{ijm}{abe} 
t^{cd}_{ml}
t^e_k
\nonumber
\end{eqnarray}

\caption{Explicit expressions for the selected three- and four-body matrix elements of the similarity-transformed form of the
the residual three-body interaction term $W_{N}$, designated by $\overline{W_{N}}^{\rm (CCSD)}$ and defined by Eq.~(\ref{sim-WN-ccsd}),
which are needed to construct the $\Xi_{a}^{i}(W_{N})$ and $\Xi_{ab}^{ij}(W_{N})$ contributions to the $\Lambda$CCSD equations,
Eqs.~(\ref{xi-ia-WN}) and~(\ref{xi-ijab-WN}), respectively.}
\label{tab:EffectiveHamiltonian3B}
\end{table}

%
%
\subsubsection{The $\Lambda$CCSD(T)-type correction for three-body Hamiltonians}
\label{sec2C-2}

We end the present section by deriving the expressions that are used in this work to determine the non-iterative
correction $\delta E^{\rm (T)}$ to the CCSD energy capable of capturing the dominant $T_{3}$ effects
in the presence of three-body interactions in the Hamiltonian. As pointed out above,
the triples correction $\delta E^{\rm (T)}$ developed in this work is an extension to 3$N$ interactions of the $\Lambda$CCSD(T)
approach, formulated for two-body Hamiltonians in Refs.~\cite{ref:26,bartlett2008a}.
We begin, however, with the more general CR-CC(2,3) methodology,
originally introduced in Refs.~\cite{crccl_jcp,crccl_cpl} and examined in the nuclear context
in Refs.~\cite{nuclei10,nuclei11}, which contains all kinds of
non-iterative triples corrections to CCSD, including $\Lambda$CCSD(T), as approximations.
The CR-CC(2,3) expressions provide us with a transparent mechanism for identifying the
additional terms in the $\Lambda$CCSD(T)-type equations that originate from the explicit
inclusion of the 3$N$ interactions in the Hamiltonian.

In general, the CR-CC(2,3), CR-CC(2,4), and other approaches in the so-called CR-CC($m$,$m^{\prime}$) 
hierarchy~\cite{crccl_jcp,crccl_cpl,crccl_jpc,crccl_ijqc2,jspp-chemphys2012}, and various
closely related approximations, including
CCSD[T]~\cite{ref:24a,ref:20},
CCSD(T)~\cite{ccsdpt},
CCSD(TQ$_{\rm f})$~\cite{ccsdtq-f},
$\Lambda$CCSD(T)~\cite{ref:26,bartlett2008a},
$\Lambda$CCSD(TQ$_{\rm f}$)~\cite{lambda-ccsd-tq},
CCSD(2)$_{\rm T}$~\cite{gwaltney1,gwaltney3,eomccpt,ccsdpt2},
CCSD(2)~\cite{gwaltney1,gwaltney3,eomccpt,ccsdpt2},
CR-CCSD(T)~\cite{leszcz,ren1,ren2,irpc,PP:TCA},
CR-CCSD(TQ)~\cite{leszcz,ren1,ren2,irpc,PP:TCA},
CR-CC(2,3)+Q~\cite{msg65},
LR-CCSD(T)~\cite{ndcmmcc},
and
LR-CCSD(TQ)~\cite{ndcmmcc},
are based on the idea of adding {\it a posteriori}, non-iterative corrections due to the higher-order cluster components,
such as $T_{3}$ or $T_{4}$, to the energies resulting from the CCSD (or some other lower-level CC) calculations.
One of the most convenient approaches for deriving these corrections is by examining the CC energy functional,
which is defined as (see, e.g., Refs.~\cite{eomcc3,lrcc4,ecc1,ecc2,ccgrad,xcc4,moszynski} and Eqs.~(\ref{eq:cc}) and~(\ref{bracc});
cf., also, Refs. \cite{gauss,bartlett-musial2007,rod-gradient,pp_rjb,jspp-chemphys2012} for reviews)
\beq
\Delta E = \langle \tilde{\Psi}| H_{N} | \Psi \rangle = \langle \Phi | (1 + \Lambda) \overline{H_{N}} | \Phi \rangle,
\eeq
or, more precisely, its asymmetric analog, which in the case of correcting the CCSD energy can be written as~\cite{crccl_jcp,crccl_cpl,jspp-chemphys2012}
\beq
\Delta E =  \langle \Phi | {\mathscr L} \, \overline{H_{N}}^{\rm (CCSD)} | \Phi \rangle ,
\label{asymmetric-energy}
\eeq
where $\overline{H_{N}}^{\rm (CCSD)}$ is the similarity-transformed Hamiltonian of CCSD, Eq.~(\ref{hbarccsd}).
The usefulness of the above expression in the context of correcting the CCSD results for the effects of
higher--than--doubly excited clusters
stems from the fact that Eq.~(\ref{asymmetric-energy}) is equivalent to the exact (i.e., full CI)
correlation energy when $\langle \Phi | {\mathscr L}$ represents the lowest-energy
left eigenstate of $\overline{H_{N}}^{\rm (CCSD)}$ obtained by diagonalizing
the latter operator in the entire $A$-particle Hilbert space. Indeed,
when the hole-particle deexcitation operator ${\mathscr L}$
entering Eq.~(\ref{asymmetric-energy}) originates from
parametrizing the full CI bra state through the ansatz
$\langle \Psi | \sim \langle \Phi | {\mathscr L} e^{-T^{\rm (CCSD)}}$,
where we assume the normalization condition
$\langle \Phi | {\mathscr L} | \Phi \rangle = 1$, the asymmetric energy expression given by Eq.~(\ref{asymmetric-energy})
produces the exact correlation energy. At the same time, since the matrix elements
$\langle\Phi_{i}^{a}| {\overline{H_{N}}}^{\rm (CCSD)}|\Phi\rangle$ and
$\langle\Phi_{ij}^{ab}| {\overline{H_{N}}}^{\rm (CCSD)}|\Phi\rangle$ vanish in the CCSD case
as required by Eqs.~(\ref{ccsd:eqt1}) and (\ref{ccsd:eqt2}),
it is easy to demonstrate that the lowest-energy
eigenvalue of $\overline{H_{N}}^{\rm (CCSD)}$ in the subspace of the Hilbert space
spanned by the reference determinant $|\Phi\rangle$ and the singly and doubly excited
determinants $|\Phi_{i}^{a}\rangle$ and $|\Phi_{ij}^{ab}\rangle$ is the CCSD correlation energy
$\Delta E^{\rm (CCSD)}$. Thus,
as shown for example in Refs.~\cite{crccl_jcp,crccl_cpl,gwaltney1,gwaltney3,eomccpt}
(cf. Ref.~\cite{jspp-chemphys2012} for a review), we can formally
split the exact correlation energy $\Delta E$ into the CCSD part
$\Delta E^{\rm (CCSD)}$ and the non-iterative correction $\delta E$
that describes all of the remaining correlations missing in CCSD
by inserting the resolution of the identity in the $A$-particle Hilbert space, written as
\beq
| \Phi \rangle \langle \Phi | + P + Q = 1,
\label{resolution-of-identity}
\eeq
where
\beq
P = P_{1} + P_{2},
\label{proj-p-def}
\eeq
\beq
Q = P_{3} + \cdots + P_{A},
\label{prof-q-def}
\eeq
and
\beq
P_{n} =  \sum_{i_{1} < \cdots < i_{n} \atop a_{1} < \cdots < a_{n}}
| \Phi_{i_{1} \ldots i_{n}}^{a_{1} \ldots a_{n}} \rangle \langle
\Phi_{i_{1} \ldots i_{n}}^{a_{1} \ldots a_{n}} |,
\label{proj-pn-def}
\eeq
into Eq.~(\ref{asymmetric-energy}), and
perform some additional manipulations that lead to
\beq
\Delta E = \Delta E^{\rm (CCSD)} +  \langle \Phi | {\mathscr L}  Q \, \overline{H_{N}}^{\rm (CCSD)} | \Phi \rangle .
\label{deltaE_LQ}
\eeq 
The resulting biorthogonal moment expansions of $\delta E$, 
which result in the aforementioned CR-CC($m$,$m^{\prime}$) 
hierarchy~\cite{crccl_jcp,crccl_cpl,crccl_jpc,crccl_ijqc2,jspp-chemphys2012},
or the perturbative expansions of $\delta E$ employing  L{\" o}wdin's partitioning technique~\cite{lowdin},
as in Refs.~\cite{gwaltney1,gwaltney3,eomccpt,ccsdpt2,ref:26,bartlett2008a} (cf., also, Ref.~\cite{stanton1997}),
which lead to methods such as $\Lambda$CCSD(T), $\Lambda$CCSD(TQ$_{\rm f}$) or CCSD(2),
provide us with the desired mathematical expressions for the non-iterative corrections due to
$T_{3}$, $T_{4}$, and other higher-order clusters.

In particular, the leading post-CCSD term in the difference $\delta E$ between the exact and CCSD
energies, which emerges from the above considerations and which captures the correlation effects due to
the connected $T_{3}$ clusters can be represented by the
following generic form~\cite{crccl_jcp,crccl_cpl,jspp-chemphys2012}
\beq
\delta E^{\rm (T)} = \langle \Phi | {\mathscr L}_{3} \, \overline{H_{N}}^{\rm (CCSD)} | \Phi \rangle
= \tsix \, {\ell}_{abc}^{ijk} \; {\mathfrak M}^{abc}_{ijk} ,
\label{delta23l}
\eeq
where
\beq
{\mathscr L}_{3} = \tsix \, \sum_{i,j,k,a,b,c} {\ell}_{abc}^{ijk} \; 
a_{a}^{\dagger} a_{b}^{\dagger} a_{c}^{\dagger} a_{k} a_{j} a_{i}
\label{ell3-def}
\eeq
is the three-body component of the exact ${\mathscr L}$ operator entering
Eq.~(\ref{asymmetric-energy}) and (\ref{deltaE_LQ}), with ${\ell}_{abc}^{ijk}$ representing
the corresponding matrix elements, and
\beq
{\mathfrak M}^{abc}_{ijk} = \langle \Phi_{ijk}^{abc}| \overline{H_{N}}^{\rm (CCSD)} | \Phi \rangle
= \langle \Phi_{ijk}^{abc}| (\overline{H_{N}}^{\rm (CCSD)})_{\rm open} | \Phi \rangle
\label{mom3}
\eeq
are the so-called generalized moments of the CCSD equations~\cite{moments,leszcz,ren1,ren2,irpc,PP:TCA}
corresponding to projections of these equations on the
triply excited determinants. At this point, the above expressions are still exact, i.e., one would have to diagonalize
$\overline{H_{N}}^{\rm (CCSD)}$ in the entire $A$-particle Hilbert space to extract the ${\mathscr L}_{3}$
component of ${\mathscr L}$ that enters Eq.~(\ref{delta23l}). Thus, in order to apply Eq.~(\ref{delta23l}) in practice, we have to
develop practical recipes for determining ${\mathscr L}_{3}$ or ${\ell}_{abc}^{ijk}$ that rely
on the information that one can extract from CCSD-level calculations. The CR-CC(2,3) approach
of Refs.~\cite{crccl_jcp,crccl_cpl} and the $\Lambda$CCSD(T) method of Refs.~\cite{ref:26,bartlett2008a},
in which some higher-order terms in the CR-CC(2,3) expressions for the $\delta E^{\rm (T)}$
correction are neglected, provide such recipes.

In the CR-CC(2,3) theory of Refs.~\cite{crccl_jcp,crccl_cpl}, presented here in the general,
orbital-rotation invariant form, where
in analogy to the CCSD energy, the resulting triples correction $\delta E^{\rm (T)}$ is invariant with respect
to rotations among the
occupied and unoccupied single-particle states, we determine the desired ${\mathscr L}_{3}$ operator or the corresponding
amplitudes ${\ell}_{abc}^{ijk}$, which enter
Eq.~(\ref{delta23l}), in a quasi-perturbative manner, using the expression (see~\cite{crccl_jcp,crccl_cpl,jspp-chemphys2012})
\beq
\langle \Phi | {\mathscr L}_{3} = \langle \Phi | (1 + \Lambda^{\rm (CCSD)}) \, \overline{H_{N}}^{\rm (CCSD)} {\mathscr R}_{3}^{\rm (CCSD)} ,
\label{ell3-crcc23}
\eeq
where
\beq
{\mathscr R}_{3}^{\rm (CCSD)} = \frac{P_{3}}{\Delta E^{\rm (CCSD)} - \overline{H_{N}}^{\rm (CCSD)}} ,
\label{resolvent}
\eeq
with
\beq
P_{3} = \sum_{i<j<k \atop a<b<c} |\Phi_{ijk}^{abc}\rangle \langle \Phi_{ijk}^{abc}| ,
\label{proj3}
\eeq
is the appropriate reduced resolvent of $\overline{H_{N}}^{\rm (CCSD)}$ in the subspace spanned by
the triply excited determinants $|\Phi_{ijk}^{abc}\rangle$ and $\Lambda^{\rm (CCSD)}$ is
the familiar $\Lambda$ operator obtained by solving the left-eigenstate CCSD equations,
Eqs.~(\ref{ccsd:eqlambda1}) and~(\ref{ccsd:eqlambda2}).
As a result, the CR-CC(2,3) correction $\delta E^{\rm (T)}$, which offers an accurate
representation of the $T_{3}$ effects on the correlation energy without forcing one to solve
for $T_{3}$ using the full CCSDT approach, assumes the following compact form:
\beq
\delta E^{\rm (T)} = \langle \Phi | (1 + \Lambda^{\rm (CCSD)}) \, \overline{H_{N}}^{\rm (CCSD)}
{\mathscr R}_{3}^{\rm (CCSD)} \, \overline{H_{N}}^{\rm (CCSD)} | \Phi \rangle .
\label{crcc23-corr}
\eeq
Alternatively, to avoid the explicit construction of the reduced resolvent ${\mathscr R}_{3}^{\rm (CCSD)}$,
Eq.~(\ref{resolvent}), in the above expression for $\delta E^{\rm (T)}$,
we can determine the ${\ell}_{abc}^{ijk}$ amplitudes by solving the linear system
\begin{eqnarray}
\sum_{l < m < n \atop d < e < f}
& &
\!\!\!\!\!\!\!\!
\langle \Phi_{lmn}^{def} | (\Delta E^{\rm (CCSD)} - \overline{H_{N}}^{\rm (CCSD)}) |\Phi_{ijk}^{abc}\rangle \:
{\ell}^{lmn}_{def}
\nonumber \\
& = & \langle \Phi | (1 + \Lambda^{\rm (CCSD)}) \, \overline{H_{N}}^{\rm (CCSD)} |\Phi_{ijk}^{abc}\rangle ,
\label{lefteqtriples}
\end{eqnarray}
which can be further simplified to
\begin{eqnarray}
-
\sum_{l < m < n \atop d < e < f}
& &
\!\!\!\!\!\!\!\!\!\!
\langle \Phi_{lmn}^{def} | (\overline{H_{N}}^{\rm (CCSD)})_{\rm open} |\Phi_{ijk}^{abc}\rangle \:
{\ell}^{lmn}_{def}
\nonumber \\
& = &
\langle \Phi | (1 + \Lambda^{\rm (CCSD)}) \, (\overline{H_{N}}^{\rm (CCSD)})_{\rm open} |\Phi_{ijk}^{abc}\rangle ,
\label{lefteqtriples-alt}
\end{eqnarray}
and use the resulting values of ${\ell}_{abc}^{ijk}$, along with the generalized moments
${\mathfrak M}^{abc}_{ijk}$, Eq.~(\ref{mom3}), to calculate $\delta E^{\rm (T)}$.
As explained in Refs.~\cite{crccl_jcp,crccl_cpl,jspp-chemphys2012}, we obtain
Eq.~(\ref{ell3-crcc23}), or the equivalent linear system given by Eq.~(\ref{lefteqtriples}),
by approximating the exact ${\mathscr L}$ operator in the left eigenvalue problem
$\langle \Phi | {\mathscr L} \, \overline{H_{N}}^{\rm (CCSD)} = \Delta E \, \langle \Phi | {\mathscr L}$,
which this operator has to satisfy and
which we right-project on the triply excited determinants $|\Phi_{ijk}^{abc}\rangle$,
by the sum of $(1 + \Lambda^{\rm (CCSD)})$, obtained by solving the left-eigenstate CCSD equations,
Eqs.~(\ref{ccsd:eqlambda1}) and (\ref{ccsd:eqlambda2}), and the unknown ${\mathscr L}_{3}$ component,
and by replacing the exact correlation energy $\Delta E$ in the resulting equations by
its CCSD counterpart $\Delta E^{\rm (CCSD)}$.

The above is the most general form of the CR-CC(2,3) theory, which encompasses other forms of
 non-iterative triples corrections available in the literature, such as $\Lambda$CCSD(T), and which satisfies a number of important
properties, including the aforementioned rotational invariance (mischaracterized
in Ref.~\cite{bartlett2008a}, but correctly described here) and the strict size extensivity
characterizing all of the commonly used CC approaches, such as CCSD
or CCSDT. If we are willing to lift the requirement of the strict invariance of the
$\delta E^{\rm (T)}$ correction with respect to arbitrary rotations among the occupied and unoccupied orbitals,
which can be justified by the fact that typical calculations of such corrections, including those
presented in this work, utilize the Hartree-Fock (i.e., fixed) orbitals, we can
eliminate the iterative steps associated with the need for solving
the linear system for the ${\ell}_{abc}^{ijk}$ amplitudes, Eq.~(\ref{lefteqtriples}) 
or~(\ref{lefteqtriples-alt}), and replace those steps by non-iterative expressions, such 
as~\cite{crccl_jcp,crccl_cpl,crccl_jpc,crccl_ijqc2,jspp-chemphys2012}
\beq
{\ell}_{abc}^{ijk} =
\langle \Phi | (1 + \Lambda^{\rm (CCSD)}) \, (\overline{H_{N}}^{\rm (CCSD)})_{\rm open} |\Phi_{ijk}^{abc}\rangle /
D_{ijk}^{abc} ,
\label{l3def-diag}
\eeq
where
\begin{eqnarray}
D_{ijk}^{abc} & = & \Delta E^{\rm (CCSD)} - \langle \Phi_{ijk}^{abc} | \overline{H_{N}}^{\rm (CCSD)} |\Phi_{ijk}^{abc}\rangle
\nonumber \\
& = & - \sum_{n=1}^{3} \langle \Phi_{ijk}^{abc} | \overline{H}_{n} |\Phi_{ijk}^{abc}\rangle ,
\label{denominator}
\end{eqnarray}
if there are no degeneracies among orbitals $i$, $j$, $k$ or $a$, $b$, $c$,
with $\overline{H}_{n}$ representing the $n$-body component of $ \overline{H_{N}}^{\rm (CCSD)}$
(we still have to solve small linear subsystems of the type of
Eqs.~(\ref{lefteqtriples}) or~(\ref{lefteqtriples-alt}) for the subsets of the ${\ell}_{abc}^{ijk}$ amplitudes
involving orbital degeneracies to retain the invariance of $\delta E^{\rm (T)}$ with respect to the rotations among
degenerate orbitals, but this is much less expensive than dealing with the complete~(\ref{lefteqtriples}) or~(\ref{lefteqtriples-alt})
system). We refer the reader to 
Refs.~\cite{crccl_jcp,crccl_cpl,crccl_jpc,crccl_ijqc2,jspp-chemphys2012} for a thorough discussion of such expressions.
Encouraged by the superb performance of the CR-CC(2,3) approach in the nuclear applications
involving two-body Hamiltonians, which we reported in Refs.~\cite{nuclei10,nuclei11},
one of our future objectives is to implement the complete CR-CC(2,3) theory,
as summarized above, for Hamiltonians including 3$N$ interactions, but
in this study we focus on the simplifications in the CR-CC(2,3) expressions for the $\delta E^{\rm (T)}$ 
corrections offered by the $\Lambda$CCSD(T) approach of Refs.~\cite{ref:26,bartlett2008a}, which facilitate
the derivations of the programmable expressions for the triples correction $\delta E^{\rm (T)}$.
Considering, however, the fact that the original
publications on the $\Lambda$CCSD(T) method~\cite{ref:26,bartlett2008a} make explicit use of the assumption that the underlying
interactions in the Hamiltonian are two-body, we use the more general CR-CC(2,3) formulas,
Eqs.~(\ref{delta23l})--(\ref{denominator}),
to identify terms in the $\Lambda$CCSD(T) equations for $\delta E^{\rm (T)}$ that result from adding the
3$N$ interactions to the Hamiltonian.

The $\Lambda$CCSD(T) approach is formally obtained by keeping only the lowest-order terms in the definitions of
the moments ${\mathfrak M}^{abc}_{ijk}$, Eq.~(\ref{mom3}), and amplitudes ${\ell}^{ijk}_{abc}$,
Eqs.~(\ref{lefteqtriples}),~(\ref{lefteqtriples-alt}), or~(\ref{l3def-diag}), that define the CR-CC(2,3)
correction $\delta E^{\rm (T)}$. Thus, assuming that the Hamiltonian
contains up to three-body interactions,
we approximate the moments ${\mathfrak M}^{abc}_{ijk}$, Eq.~(\ref{mom3}), by retaining terms
in $(\overline{H_{N}}^{\rm (CCSD)})_{\rm open}$
that are at most linear in $T$, i.e.,
\begin{eqnarray}
{\mathfrak M}^{abc}_{ijk} & \approx & \langle \Phi_{ijk}^{abc}| [H_{N} (1 + T_{1} + T_{2})]_{C} | \Phi \rangle
\nonumber \\
& = &
{\mathfrak M}^{abc}_{ijk}({\rm 2B}) + {\mathfrak M}^{abc}_{ijk}(W_{N}) ,
\label{Mijkabc}
\end{eqnarray}
where the NO2B contribution to ${\mathfrak M}^{abc}_{ijk}$ is given by
\begin{eqnarray}
{\mathfrak M}^{abc}_{ijk}({\rm 2B}) & = &
\langle \Phi_{ijk}^{abc}| [H_{N,{\rm 2B}} (1 + T_{1} + T_{2})]_{C} | \Phi \rangle
\nonumber \\
& = &
\langle \Phi_{ijk}^{abc}| (V_{N} T_{2})_{C} | \Phi \rangle
\label{Mijkabc-NO2B}
\end{eqnarray}
and the contribution due to the residual 3$N$ interactions has the form
\beq
{\mathfrak M}^{abc}_{ijk}(W_{N}) = \langle \Phi_{ijk}^{abc}| [W_{N} (1 + T_{1} + T_{2})]_{C} | \Phi \rangle .
\label{Mijkabc-WN}
\eeq
In order to derive the analogous expressions for the amplitudes ${\ell}^{ijk}_{abc}$, which would be
consistent with the approximations that lead to the non-iterative $\Lambda$CCSD(T) approach
of Refs.~\cite{ref:26,bartlett2008a}, where one makes an assumption that the Fock operator
is diagonal in the occupied and unoccupied single-particle spaces,
so that $f_{j}^{i} = \epsilon_{i} \delta_{ij}$ and $f_{b}^{a} = \epsilon_{a} \delta_{ab}$,
where $\epsilon_p$ represents the diagonal matrix element $f_{p}^{p}$,
which is automatically
satisfied by the calculations reported in this study since they rely on the canonical Hartree-Fock orbitals,
we replace
the reduced resolvent ${\mathscr R}_{3}^{\rm (CCSD)}$ entering the CR-CC(2,3) correction
$\delta E^{\rm (T)}$, Eq.~(\ref{crcc23-corr}), by its simplified M{\o}ller-Plesset form
adopted in the $\Lambda$CCSD(T) considerations~\cite{ref:26,bartlett2008a}, i.e.,
\begin{eqnarray}
{\mathscr R}_{3}^{\rm (CCSD)} & = & - \frac{P_{3}}{(\overline{H_{N}}^{\rm (CCSD)})_{\rm open}}
\approx
- \frac{P_{3}}{F_{N}}
\nonumber \\
& = & \sum_{i<j<k \atop a<b<c} (\epsilon^{abc}_{ijk})^{-1} |\Phi_{ijk}^{abc} \rangle \langle \Phi_{ijk}^{abc}| ,
\label{resolvent-approx}
\end{eqnarray}
where
\beq
\epsilon_{ijk}^{abc} = \epsilon_i + \epsilon_j + \epsilon_k - \epsilon_a - \epsilon_b - \epsilon_c
\label{epsilon}
\eeq
is the orbital energy difference for triples.
The latter approximation is equivalent to replacing $(\overline{H_{N}}^{\rm (CCSD)})_{\rm open}$ on the left-hand side
of the linear system given by Eq.~(\ref{lefteqtriples-alt}), which corresponds 
to the more elaborate CR-CC(2,3) treatment, by the $F_{N}$ operator. If we further approximate $\overline{H_{N}}^{\rm (CCSD)}$ on the
right-hand side of Eq.~(\ref{lefteqtriples-alt}) by the leading contribution to $\overline{H_{N}}^{\rm (CCSD)}$, which is
the normal-ordered Hamiltonian $H_{N}$ itself, we can replace the linear system given by Eq.~(\ref{lefteqtriples-alt}) by its simplified form
\begin{eqnarray}
\lefteqn{
- 
\sum_{l < m < n \atop d < e < f}
\langle \Phi_{lmn}^{def} | F_{N} |\Phi_{ijk}^{abc}\rangle \:
{\ell}^{lmn}_{def}
}
\nonumber \\
&&
\ \ \ \ \ = \langle \Phi | (1 + \Lambda^{\rm (CCSD)}) \, H_{N} |\Phi_{ijk}^{abc}\rangle \ ,
\label{lefteqtriples-sim}
\end{eqnarray}
which immediately allows us to write
\beq
{\ell}_{abc}^{ijk} = (\epsilon^{abc}_{ijk})^{-1} \langle \Phi | (1 + \Lambda^{\rm (CCSD)}) \, H_{N} |\Phi_{ijk}^{abc}\rangle .
\label{ell-lambdacc}
\eeq
After splitting the above expression into the NO2B and residual 3$N$ contributions and identifying the
non-vanishing terms, we obtain
\beq
{\ell}_{abc}^{ijk} = {\ell}_{abc}^{ijk}({\rm 2B}) + {\ell}_{abc}^{ijk}(W_{N}) ,
\label{ell-abc-ijk}
\eeq
where
\begin{eqnarray}
{\ell}_{abc}^{ijk}({\rm 2B}) & = & \langle \Phi | [(\Lambda_{1} V_{N})_{DC} + (\Lambda_{2} F_{N})_{DC}
\nonumber \\
&&
+ (\Lambda_{2} V_{N})_{C}] |\Phi_{ijk}^{abc} \rangle / \epsilon^{abc}_{ijk}
\label{numerator-NO2B}
\end{eqnarray}
and
\begin{eqnarray}
{\ell}_{abc}^{ijk}(W_{N}) & = & \langle \Phi | [W_{N} + (\Lambda_{1} W_{N})_{C}
\nonumber \\
&&
+ (\Lambda_{2} W_{N})_{C}] |\Phi_{ijk}^{abc} \rangle / \epsilon^{abc}_{ijk} .
\label{numerator-WN}
\end{eqnarray}
Equation~(\ref{delta23l}), with moments ${\mathfrak M}^{abc}_{ijk}$ approximated by
Eqs.~(\ref{Mijkabc})--(\ref{Mijkabc-WN}) and amplitudes ${\ell}_{abc}^{ijk}$ by
Eqs.~(\ref{ell-abc-ijk})--(\ref{numerator-WN}), is the desired extension of the $\Lambda$CCSD(T)
correction due to the connected $T_{3}$ clusters to the 3$N$ interaction case.
By comparing the expressions for the NO2B contributions to ${\mathfrak M}^{abc}_{ijk}$
and ${\ell}_{abc}^{ijk}$ given by Eqs.~(\ref{Mijkabc-NO2B}) and~(\ref{numerator-NO2B}), respectively, with
the analogous formulas for the two-body Hamiltonians reported in Ref.~\cite{bartlett2008a}, we can
immediately see that the $\Lambda$CCSD(T) approach presented here, which we derived by simplifying the
CR-CC(2,3) equations, reduces to the $\Lambda$CCSD(T) theory of Refs.~\cite{ref:26,bartlett2008a},
when the Hamiltonian of interest contains pairwise interactions only.

Based on the above considerations, we can give the triples correction formula for three-body Hamiltonians,
within the $\Lambda$CCSD(T) approximation scheme discussed in this work, the physically meaningful form
\beq
\delta E^{\rm (T)} = \delta E_{\rm 2B}^{\rm (T)} + \delta E_{\rm 3B}^{\rm (T)} ,
\label{triples-energy}
\eeq
where the pure NO2B contribution $\delta E_{\rm 2B}^{\rm (T)}$ is defined as
\beq
\delta E_{\rm 2B}^{\rm (T)} =
\tsix \, {\ell}_{abc}^{ijk}({\rm 2B}) \: {\mathfrak M}_{ijk}^{abc}({\rm 2B}) ,
\label{triples-2B}
\eeq
whereas the $\delta E_{\rm 3B}^{\rm (T)}$ component of $\delta E^{\rm (T)}$,
which is present only when the residual 3$N$ interactions are taken into
account, is given by
\begin{eqnarray}
\delta E_{\rm 3B}^{\rm (T)} & = &
\tsix \, [
{\ell}_{abc}^{ijk}({\rm 2B}) \: {\mathfrak M}_{ijk}^{abc}(W_{N})
+ {\ell}_{abc}^{ijk}(W_{N}) \: {\mathfrak M}_{ijk}^{abc}({\rm 2B})
\nonumber \\
&&
+ {\ell}_{abc}^{ijk}(W_{N}) \: {\mathfrak M}_{ijk}^{abc}(W_{N}) ].
\label{triples-3B}
\end{eqnarray}
The explicit $m$-scheme-type expressions for the NO2B contributions to moments ${\mathfrak M}_{ijk}^{abc}$ and
amplitudes ${\ell}_{abc}^{ijk}$, within the
$\Lambda$CCSD(T) approximation defined by 
Eqs.~(\ref{Mijkabc-NO2B}),~(\ref{Mijkabc-WN}),~(\ref{numerator-NO2B}) and~(\ref{numerator-WN}), are
\beq
{\mathfrak M}_{ijk}^{abc}({\rm 2B}) =
{\mathscr A}^{ab/c}
{\mathscr A}_{ij/k}
(
v^{ab}_{kd}
t^{dc}_{ij}
-
v^{cl}_{ij}
t^{ab}_{lk}
)
\label{M-ijkabc-2B}
\eeq
and
\begin{eqnarray}
{\ell}_{abc}^{ijk}({\rm 2B}) & = &
{\mathscr A}_{ab/c}
{\mathscr A}^{ij/k}
(
f^k_c
\lambda^{ij}_{ab}
+
v^{ij}_{ab}
\lambda^k_c
+
v^{kd}_{ab}
\lambda^{ij}_{dc}
\nonumber \\
&&
-
v^{ij}_{cl}
\lambda^{lk}_{ab}
)
/\epsilon^{abc}_{ijk} ,
\label{ell-ijkabc-2B}
\end{eqnarray}
respectively (the analogous equations can also be found in Ref.~\cite{bartlett2008a},
although the equation in Ref.~\cite{bartlett2008a}, which would be equivalent to
our Eq.~(\ref{ell-ijkabc-2B}), is applicable to real orbitals only).
For the residual 3$N$ contributions to ${\mathfrak M}_{ijk}^{abc}$ and
amplitudes ${\ell}_{abc}^{ijk}$, we can write
\begin{eqnarray}
{\mathfrak M}_{ijk}^{abc}(W_{N}) & = &
\mWME{abc}{ijk}
-
{\mathscr A}^{ab/c}
\mWME{abl}{ijk} 
t^c_l 
+
{\mathscr A}_{ij/k}
\mWME{abc}{ijd} 
t^d_k 
\nonumber \\
&&
+
\tfrac{1}{2} 
{\mathscr A}_{ij/k}
\mWME{abc}{dek} 
t^{de}_{ij} 
+
\tfrac{1}{2} 
{\mathscr A}^{ab/c}
\mWME{lmc}{ijk} 
t^{ab}_{lm} 
\nonumber \\
&&
+
{\mathscr A}^{ab/c}
{\mathscr A}_{ij/k}
\mWME{abl}{ijd} 
t^{cd}_{kl}
\label{M-ijkabc-WN}
\end{eqnarray}
and
\begin{eqnarray}
{\ell}_{abc}^{ijk}(W_{N}) & = &
(
\mWME{ijk}{abc} 
-
{\mathscr A}_{ab/c}
\mWME{ijk}{abl} 
\lambda^l_c 
+
{\mathscr A}^{ij/k}
\mWME{ijd}{abc} 
\lambda^k_d 
\nonumber \\
&&
+
\tfrac{1}{2} 
{\mathscr A}^{ij/k}
\mWME{dek}{abc} 
\lambda^{ij}_{de} 
+
\tfrac{1}{2} 
{\mathscr A}_{ab/c}
\mWME{ijk}{lmc} 
\lambda^{lm}_{ab} 
\nonumber \\
&&
+
{\mathscr A}_{ab/c}
{\mathscr A}^{ij/k}
\mWME{ijd}{abl} 
\lambda^{kl}_{cd}
)
/\epsilon^{abc}_{ijk} ,
\label{ell-ijkabc-WN}
\end{eqnarray}
respectively.
The three-index antisymmetrizers
${\mathscr A}_{pq/r}={\mathscr A}^{pq/r}$, which enter the above formulas along
with the previously defined two-index antisymmetrizers ${\mathscr A}_{pq}={\mathscr A}^{pq}$,
Eq.~(\ref{antisymm-pq}), are defined in a usual way, {\it viz.},
\beq
{\mathscr A}_{pq/r} \equiv {\mathscr A}^{pq/r} = 1 - (pr) - (qr),
\eeq
where we use the $(pq)$ symbol once again to represent a transposition of two indices.
As in the case of the CCSD and $\Lambda$CCSD equations discussed in Sec.~\ref{sec2C-1},
the $m$-scheme-style expressions represented by Eqs.~(\ref{M-ijkabc-2B})--(\ref{ell-ijkabc-WN}) can again be converted into an angular-momentum-coupled form which greatly facilitates the computations.

\bigskip

We finalize our formal presentation of the $\Lambda$CCSD(T) theory for
 three-body Hamiltonians 
by emphasizing the differences between $\Lambda$CCSD(T) in the NO2B
approximation
and the complete $\Lambda$CCSD(T) treatment including the residual
3$N$ interactions $W_{N}$.
According to the above analysis, in the full treatment of three-body
interactions within the $\Lambda$CCSD(T) description, one determines
the total energy $E$, designated as $E^{(\rm {\Lambda}CCSD(T))}$, as follows:
\begin{eqnarray}
\lefteqn{
E^{\rm ({\Lambda}CCSD(T))} = }
\nonumber \\
&&
 \ERef + \Delta E_{\rm 2B}^{\rm (CCSD)}
+ \delta E_{\rm 2B}^{\rm (T)}
+ \Delta E_{\rm 3B}^{\rm (CCSD)}
+ \delta E_{\rm 3B}^{\rm (T)} ,
\label{lambda-ccsd-t-full}
\end{eqnarray}
where we calculate the NO2B-type correlation energy contributions $\Delta E_{\rm 2B}^{\rm (CCSD)}$ and
$\delta E_{\rm 2B}^{\rm (T)}$ using Eqs.~(\ref{energy-NO2B}) and (\ref{triples-2B}), respectively,
and the contributions associated with the presence of the residual 3$N$ interactions,
$\Delta E_{\rm 3B}^{\rm (CCSD)}$ and $\delta E_{\rm 3B}^{\rm (T)}$, using 
Eqs.~(\ref{energy-WN}) and~(\ref{triples-3B}), respectively. The reference energy
$\ERef$, which obviously does not contain any information about the residual 3$N$ effects represented by the normal-ordered operator $W_{N}$,
is calculated using Eq.~(\ref{ref-energy}).
In the case of $\Lambda$CCSD(T) calculations in the NO2B approximation, we
replace the complete energy expression given by Eq.~(\ref{lambda-ccsd-t-full}) by
its simplified form, in which the $W_{N}$-containing terms,
$\Delta E_{\rm 3B}^{\rm (CCSD)}$ and $\delta E_{\rm 3B}^{\rm (T)}$, are neglected, i.e.,
\beq
E_{\rm 2B}^{\rm ({\Lambda}CCSD(T))} = \ERef + \Delta E_{\rm 2B}^{\rm (CCSD)} 
+ \delta E_{\rm 2B}^{\rm (T)} .
\label{lambda-ccsd-t-2B}
\eeq
We stress, however, that the
differences between the complete and NO2B treatments of the 3$N$ interactions in
the $\Lambda$CCSD(T) calculations are not limited to the final energy expressions. In the
most complete $\Lambda$CCSD(T) calculations, in which the three-body interactions in the Hamiltonian
are treated fully, the singly and doubly excited cluster amplitudes, $t_{i}^{a}$ and $t_{ij}^{ab}$, and
their singly and doubly deexcited $\lambda_{a}^{i}$ and $\lambda_{ab}^{ij}$ counterparts are determined from
 CCSD and left-eigenstate CCSD calculations with all terms in the normal-ordered
three-body Hamiltonian $H_{N}$, Eq.~(\ref{normal-form}), including those that contain $W_{N}$,
properly accounted for, as in
Eqs.~(\ref{ccsd:eqt1}) and~(\ref{ccsd:eqt2}) for CCSD and~(\ref{ccsd-lambda1}) and~(\ref{ccsd-lambda2}) for $\Lambda$CCSD.
This should be contrasted with the NO2B approximation to the $\Lambda$CCSD(T) approach, in which
the $t_{i}^{a}$, $t_{ij}^{ab}$, $\lambda_{a}^{i}$, and $\lambda_{ab}^{ij}$ amplitudes, which are needed to
construct the $\Delta E_{\rm 2B}^{\rm (CCSD)}$ and $\delta E_{\rm 2B}^{\rm (T)}$ energy components
in Eq.~(\ref{lambda-ccsd-t-2B}), are obtained by solving the CCSD and left-eigenstate CCSD
equations, where the $W_{N}$-containing $\Theta_{i}^{a}(W_{N})$ and $\Theta_{ij}^{ab}(W_{N})$ terms
in the CCSD system, Eqs.~(\ref{ccsd:eqt1}) and~(\ref{ccsd:eqt2}), and the
$\Xi_{a}^{i}(W_{N})$ and $\Xi_{ab}^{ij}(W_{N})$ terms in the $\Lambda$CCSD system,
Eqs.~(\ref{ccsd-lambda1}) and~(\ref{ccsd-lambda2}), are neglected. 
Clearly, very similar remarks apply to a comparison of the complete and NO2B treatments of
the 3$N$ interactions in the underlying CCSD calculations, where the corresponding total energies are
defined as
\begin{eqnarray}
E^{\rm (CCSD)} &=& \ERef + \Delta E_{\rm 2B}^{\rm (CCSD)}
+ \Delta E_{\rm 3B}^{\rm (CCSD)} \\
&\equiv& \ERef + \Delta E^{\rm (CCSD)}
\label{ccsd-full}
\end{eqnarray}
in the former case, and
\beq
E_{\rm 2B}^{\rm (CCSD)} = \ERef + \Delta E_{\rm 2B}^{\rm (CCSD)} ,
\label{ccsd-2B}
\eeq
in the latter case. One of the interesting questions that our calculations discussed in Section~\ref{sec3} try to address is
if it is beneficial to consider an intermediate $\Lambda$CCSD(T) approximation, where the 3$N$ forces are
treated fully at the CCSD level, while using the NO2B approximation in the determination of the
$\delta E^{\rm (T)}$ triples correction, so that the full $\Lambda$CCSD(T) energy expression,
Eq.~(\ref{lambda-ccsd-t-full}), is replaced by the somewhat simpler formula
\begin{eqnarray}
\tilde{E}^{\rm ({\Lambda}CCSD(T))} & = & \ERef + \Delta E^{\rm (CCSD)} + \delta E_{\rm 2B}^{\rm (T)} .
\label{lambda-ccsd-t-partial}
\end{eqnarray}

Finally, it is worth pointing out that one of the most interesting differences between the
$\Lambda$CCSD(T) calculations with the NO2B and full treatments of the 3$N$ interactions in the Hamiltonian is
the significance of the $T_{3}$ contributions induced by the residual $W_{N}$ component.
As in conventional many-body theory based on pairwise interactions,
the NO2B approximation shifts the $T_{3}$ contribution to the second and higher orders of 
many-body perturbation theory (MBPT) in the wave function and the fourth and higher MBPT orders
in the energy, since in the absence of the $W_{N}$ component in the Hamiltonian, the lowest-order
approximation to $T_{3}$ originates from the formula (cf., e.g., Ref.~\cite{ref:20}, and references therein)
$T_{3}^{(2)} |\Phi\rangle = ({\mathscr R}_{3} V_{N} {\mathscr R}_{2} V_{N})_{C} |\Phi\rangle$,
where ${\mathscr R}_{n} = - (F_{N})^{-1} {P_{n}}$ is the $n$-body component of the MBPT reduced resolvent
(assuming, for simplicity, Hartree-Fock orbitals). The fourth- and higher-order MBPT contributions to the energy due to
the $T_{3}$ clusters originating from the pairwise interaction term $V_{N}$ in $H_{N}$ are captured by
the $\delta E_{\rm 2B}^{\rm (T)}$ correction, Eq.~(\ref{triples-2B}), which is present in any form of
the $\Lambda$CCSD(T) (or even CCSD(T) or CCSD[T]) calculations, including those in which the 3$N$ interactions are completely neglected.
The situation changes when we include the residual 3$N$ interaction term $W_{N}$ in the calculations. In this case, the  $T_{3}$ cluster component
due to $W_{N}$ shows up already in the first MBPT order in the wave function and the second MBPT order in the
energy, since one can form the connected wave function diagram with six external lines representing $T_{3}$
using the formula $T_{3}^{(1)} |\Phi\rangle = ({\mathscr R}_{3} W_{N})_{C} |\Phi\rangle$. The
corresponding second-order
MBPT contribution due to the $T_{3}$ cluster component originating from the presence of $W_{N}$ in the
Hamiltonian is captured by the $\delta E_{\rm 3B}^{\rm (T)}$ correction through the last
$\tsix \, {\ell}_{abc}^{ijk}(W_{N}) \: {\mathfrak M}_{ijk}^{abc}(W_{N})$ term in Eq.~(\ref{triples-3B}),
which, based on Eqs.~(\ref{M-ijkabc-WN}) and~(\ref{ell-ijkabc-WN}), contains
the second-order $\tsix \sum_{i,j,k,a,b,c} w_{ijk}^{abc} w_{abc}^{ijk}/\epsilon^{abc}_{ijk}$ expression as the
leading component.
It is trivial to show that the latter expression is equivalent to the vacuum diagram representing
$\langle \Phi | (W_{N} T_{3}^{(1)})_{C} | \Phi \rangle$. Clearly,
such a term cannot be captured by CCSD, even when the $W_{N}$ interactions are included in the calculations,
since CCSD ignores the $T_{3}$ contributions altogether and the CCSD correlation energy can only directly engage the
$T_{1} T_{2}$ and $\six T_{1}^{3}$ clusters, as in $ \Delta E_{\rm 3B}^{\rm (CCSD)}$, Eq.~(\ref{energy-WN}).
We would have the $\langle \Phi | (W_{N} T_{3})_{C} | \Phi \rangle$ component in the
correlation energy if we used the full CCSDT approach with the residual $W_{N}$ interactions.
It is, therefore, very encouraging to observe that the extension of the $\Lambda$CCSD(T) approach to
 three-body Hamiltonians developed in this work captures the sophisticated
$T_{3}$-cluster physics originating from the residual
3$N$ forces represented by the $W_{N}$ operator, which normally requires the full CCSDT treatment,
via the $\delta E_{\rm 3B}^{\rm (T)}$ energy component defined by Eq.~(\ref{triples-3B}).
It is useful to point out that the smallness of the residual 3$N$ interaction represented by $W_N$ relative to the pairwise $V_N$ component causes the last term in Eq.~(\ref{triples-3B}), which formally shows up in second order, to be for all practical purposes negligible.
The first and second terms in Eq.~(\ref{triples-3B}) that mix the $V_N$ and $W_N$ contributions to $\delta E_{\rm 3B}^{\rm (T)}$ are larger, dominating $\delta E_{\rm 3B}^{\rm (T)}$, but they are still quite small compared to $\delta E_{\rm 2B}^{\rm (T)}$.
Numerical examples illustrating the relative significance of $\delta E_{\rm 2B}^{\rm (T)}$ vs $\delta E_{\rm 3B}^{\rm (T)}$ contributions are discussed in the next section.

%

\section{Application to Medium-Mass Nuclei}
\label{sec3}

%
%
                                                      
\subsection{Hamiltonian and basis}
\label{sec3A}

We use the chiral $NN$ interaction at N$^3$LO~\cite{EnMa03} and a local form of the chiral 3$N$ interaction at N$^2$LO~\cite{Navr07}. The initial Hamiltonian is transformed through a similarity renormalization group (SRG) evolution at the two- and three-body level to enhance the convergence behavior of the many-body calculations. The SRG transformation represents a continuous unitary transformation parametrized by a flow parameter $\alpha$, with the initial Hamiltonian corresponding to $\alpha=0$ ~\cite{JuNa09,RoNe10,RoLa11}. 
In all calculations we use the \mbox{400 MeV} reduced-cutoff version of the chiral 3$N$ interaction as described in~\cite{RoLa11,RoBi12,RoCa13,HeBo13}.
This cutoff reduction is motivated by the observation that SRG-induced $4N$ interactions have a sizable impact on ground-state energies of medium-mass nuclei, which can be reduced efficiently by lowering the cutoff.

We will employ two types of SRG-evolved Hamiltonians. The $NN$+3$N$-full Hamiltonian starts with the initial chiral $NN$+3$N$ Hamiltonian and retains all terms up to the three-body level in the SRG evolution; the $NN$+3$N$-induced Hamiltonian omits the chiral 3$N$ interaction from the initial Hamiltonian, but keeps all induced three-body terms throughout the evolution. The three-body SRG evolution is performed in a harmonic-oscillator (HO) model space with up to 40 oscillator 
quanta~\cite{RoLa11,RoCa13}. To ensure the sufficiency of this model space for smaller HO frequencies we apply a frequency conversion technique~\cite{RoCa13}. Thus, we evolve the Hamiltonian at an adequate HO frequency, which is set here ta $\hbar\Omega=28\,$MeV, and convert the Hamiltonian matrix elements to the HO basis with the desired frequency for the many-body calculation afterwards. Furthermore, we consider a range of flow parameters $\alpha$ in order to observe how the individual contributions in the CC calculations evolve with the SRG flow.
We note that all calculations are performed with the intrinsic Hamiltonians and that no correction for spurious center-of-mass effects is applied since those are expected to be small~\cite{HaPa09}.

For our CC calculations, the underlying single-particle basis is a HO basis truncated in the principal oscillator quantum number $2n+l = e \leq e_{\max}$ and we go up to $e_{\max}=12$. We perform Hartree-Fock calculations explicitly including  the 3$N$ interaction for each set of basis parameters to obtain an optimized single-particle basis and to stabilize the convergence of the CC iterations.  
Due to their enormous number, it is not possible to include all 3$N$ matrix elements that would appear in the larger bases. Therefore, regarding computing time, we restrict our calculations to  three-body matrix elements with $e_1+e_2+e_3 \leq E_{3\max} = 12$. For this particular value of $E_{3\max}$ we capture a significant part of the 3$N$ interaction, but, mostly for the harder interactions, we are not yet fully converged with respect to $E_{3\max}$ \cite{BiLa13}. However, this is not expected to impact the discussion in this article.

For closed-shell nuclei we use an angular-momentum coupled formulation of CC theory~\cite{HaPa10} which enables us to operate with reduced matrix elements for all operators involved, in particular the Hamiltonian. This leads to a drastic reduction of the number of matrix elements to be processed compared to an $m$-scheme description and hence greatly extends the range of the method to medium-mass nuclei and beyond.

%
\subsection{Results}
\label{sec3B}

\begin{figure}[t]
\centering{ \includegraphics[width=1.0\columnwidth]{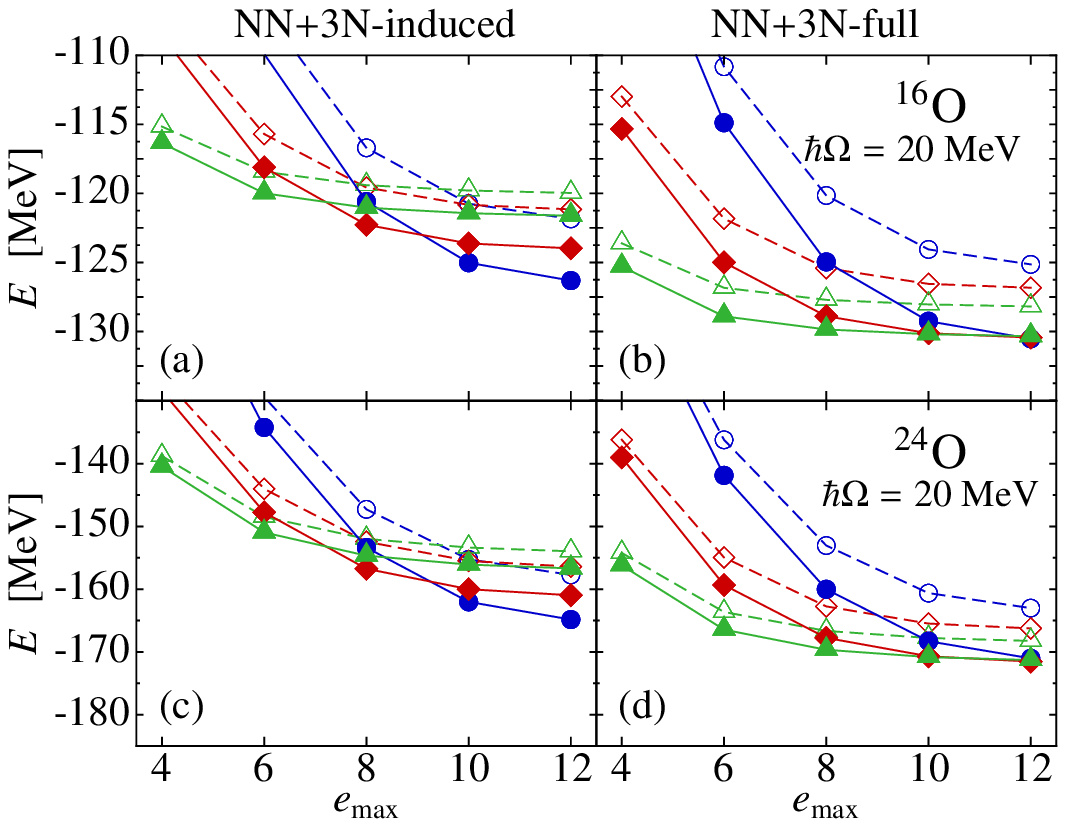}  \\[-10pt] 
}
\vspace{-5pt}
\caption{
(color online) CCSD (dashed lines) and $\Lambda$CCSD(T) (solid lines) ground-state energies for \elem{O}{16} and \elem{O}{24} as function of $e_{\max}$ for the two types of Hamiltonians (see column headings) with $E_{3\max}=12$ for SRG flow parameters 
$\alpha$ set at $0.02\,\text{fm}^4$ (\symbolcircle[FGBlue]), $0.04\,\text{fm}^4$ (\symboldiamond[FGRed]), and $0.08\,\text{fm}^4$ (\symboltriangle[FGGreen]).
}	
\label{fig:CCSDvsLCCSDT}
\end{figure}

To assess the overall importance of triply excited clusters in nuclear-structure calculations, in Fig.~\ref{fig:CCSDvsLCCSDT} we compare the CCSD and $\Lambda$CCSD(T) ground-state energies $E^{\rm (CCSD)}$ and $E^{\rm ({\Lambda}CCSD(T))}$ using the complete 3$N$ information, as function of $\emax$ for \elem{O}{16} and \elem{O}{24} and for the two 3$N$ Hamiltonians discussed in the previous section.
First, we notice that we are reasonably converged within the model spaces we operate in and we observe the expected faster convergence with respect to model space-size for the softer, further evolved, interactions. Furthermore, the triples correction $\delta E^{\rm (T)}$ provides about \mbox{2--5\,\%} of the binding energy for all nuclei considered, where, as expected, the contribution of the triply excited clusters decreases with the SRG flow parameter. 
Therefore, if one eventually aims at an accuracy in ground-state calculations of about \mbox{1\,\%},
the truncation in the cluster operator $T$ is identified as one of the larger sources of
error. The CCSD level of theory is not sufficiently accurate, the connected triply excited effects are not negligible, even for the softest interaction considered.

\begin{figure}[p]
 \includegraphics{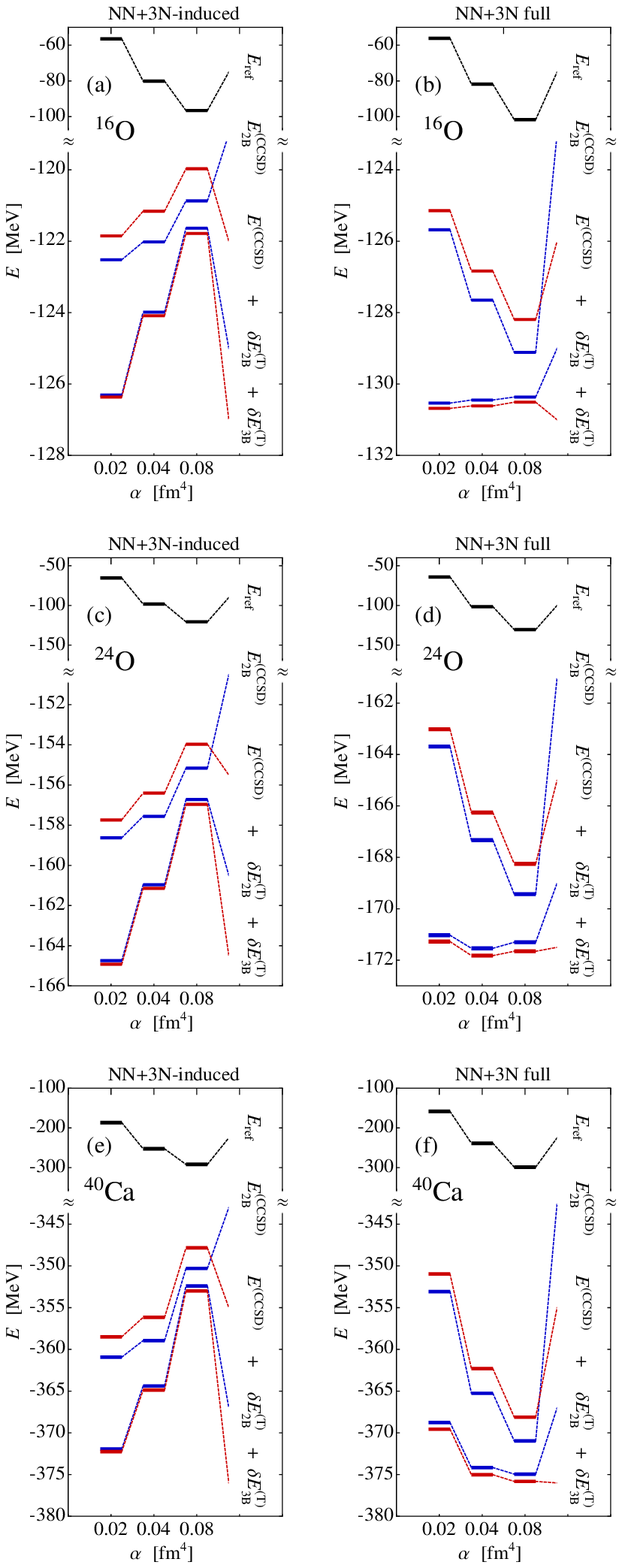}
 
 \vspace{-10pt}
\caption{
(color online) Anatomy of individual contributions from CCSD and $\Lambda$CCSD(T) to the total binding energy of \elem{O}{16,24} and \elem{Ca}{40} for the two types of Hamiltonians with $E_{3\max}=12$ and SRG flow parameters $\rm \alpha = 0.02, 0.04 \ and \ 0.08\,\text{fm}^4$. For \elem{O}{16,24}, an $\emax =12$ model space and 
oscillator frequency $\hbar \Omega = 20$\,MeV was used, whereas for \elem{Ca}{40} we work in an $\emax = 10$ model space with $\hbar \Omega = 24$\,MeV.
}	
\label{fig:IndividualContributions}
\end{figure}

Next we address the importance of the residual 3$N$ interaction in CCSD and $\Lambda$CCSD(T) calculations. Our discussion is  complicated by the fact that energy values are not only determined by their expressions in terms of the $T^{\rm(CCSD)}$ and $\Lambda^{\rm(CCSD)}$ amplitudes $t^{a}_{i}$, $t^{ab}_{ij}$ and  $\lambda^i_a$, $\lambda^{ij}_{ab}$, but also by the type of equations -- with or without inclusion of the $W_N$ terms -- used to determine the amplitudes. This leads to various possible and reasonable combinations to consider.

In Fig.~\ref{fig:IndividualContributions}, we show results for a series of increasingly complete calculations of the energy for \elem{O}{16}, \elem{O}{24}, and \elem{Ca}{40} and for both the $NN+3N$-induced and $NN+3N$-full Hamiltonians. The energy $E_{\rm 2B}^{\rm (CCSD)}$ is calculated in NO2B approximation, i.e., the $W_N$ terms are  neglected in the equations determining the $T^{\rm(CCSD)}$ amplitudes. For the calculation of all other energies we use $T^{\rm(CCSD)}$ and $\Lambda^{\rm(CCSD)}$ amplitudes determined from their respective amplitude   equations including the $W_N$ terms.
By comparing $E_{\rm 2B}^{\rm (CCSD)}$ with $E^{\rm (CCSD)}$, we obtain a direct quantification of the combined effect of the additional $W_N$ terms in the CCSD amplitude equations and energy expression. Note that $E^{\rm (CCSD)} - E_{\rm 2B}^{\rm (CCSD)} \ne \Delta E_{\rm 3B}^{\rm (CCSD)}$ here, due to the use of different amplitudes.
The interesting question of whether the $W_N$ terms are more important in the determination of the amplitudes or in the energy expression will be adressed further below.
Contrary to the previous situation, the same amplitudes are used in the calculation of $\delta E_{\rm 2B}^{\rm (T)}$ and $\delta E_{\rm 3B}^{\rm (T)}$. Therefore, using these numbers we can only quantify how important the $W_N$ contributions, given simply  by $\delta E_{\rm 3B}^{\rm (T)}$, are in the calculation of the total triples correction $\delta E^{\rm(T)}$, i.e., we can compare the approximate energy expression $\tilde{E}^{\rm ({\Lambda}CCSD(T))}$, Eq.~(\ref{lambda-ccsd-t-partial}), with the full form $E^{\rm ({\Lambda}CCSD(T))}$, Eq.~(\ref{lambda-ccsd-t-full}), but we cannot at the same time  assess the relevance of $W_N$ terms in the respective equations determining the $T^{\rm(CCSD)}$ and $\Lambda^{\rm(CCSD)}$ amplitudes. Particularly for $\delta E_{\rm 2B}^{\rm (T)}$, other choices of where to include $W_N$ terms in the amplitude equations seem reasonable. We come back to this issue below but already mention here that for $\delta E_{\rm 2B}^{\rm (T)}$ other choices of amplitude equations lead to practically the same results.

All data shown in Fig.~\ref{fig:IndividualContributions} are compiled in Table~\ref{tab:Data}, and in the following
we consider \elem{O}{16} with the $NN$+3$N$-full Hamiltonian \mbox{[Fig.~\ref{fig:IndividualContributions}(b)]} at flow parameter values $\alpha$ = 0.02\,\fm and 0.08\,\fm as an example. 
When $\alpha$ increases, more and more of the binding energy is shifted to lower orders of the cluster expansion and the contributions from the higher orders consequently get smaller with the SRG flow: the magnitude of the reference energy $\ERef$ grows from \mbox{--56.11\,MeV} to \mbox{--101.67\,MeV}, while the CCSD correlation energy $\Delta E^{\rm (CCSD)}$ decreases from \mbox{--69.03\,MeV} to \mbox{--26.52\,MeV} as we go from $\alpha$ = 0.02\,\fm to 0.08\,\fm and the $\Lambda$CCSD(T) energy correction $\delta E^{\rm (T)}$, which we also consider as a measure for the contributions of the omitted cluster operators beyond the three-body level~\cite{BiLa13}, decreases from \mbox{--5.54\,MeV} to \mbox{--2.34\,MeV}, corresponding to \mbox{4.2\,\%} and \mbox{1.8\,\%} of the total binding energy. In the medium-mass regime considered here, these uncertainties related to the cluster truncation are typically the largest in our calculations for a given Hamiltonian, and therefore they determine the overall level of accuracy we aim at~\cite{BiLa13}.

Examining the contributions from the residual 3$N$ interaction to $\Delta E^{\rm (CCSD)}$ we find that, while the absolute value of $\Delta E^{\rm (CCSD)}$ decreases by about 30 MeV when we evolve the Hamiltonian from $\alpha$ = 0.02\,\fm further to 0.08\,\fmNoSpace, $\Delta E^{\rm (CCSD)}- \Delta E_{\rm 2B}^{\rm (CCSD)}$ is only subject to a slight increase from \mbox{0.54\,MeV} to \mbox{0.92\,MeV}, corresponding to \mbox{0.4\,\%} and \mbox{0.7\,\%} of the total binding energy. 
Consequently, the relative as well as the absolute importance of the residual 3$N$ interaction to the CCSD correlation energy grows with the SRG flow.

Furthermore, while for the harder Hamiltonian at $\alpha$ = 0.02\,\fm the $W_N$ contributions to $\Delta E^{\rm (CCSD)}$ are about one order of magnitude smaller than the accuracy level set by $\delta E^{\rm (T)}$, for the softer $\alpha$ = 0.08\,\fm Hamiltonian the $W_N$ contributions have an comparable size of about \mbox{39\,\%} of the triples correction. Therefore, in order to keep different errors at a consistent level, for soft interactions the residual 3$N$ contributions should be included in CCSD if the triples correction is considered as well.

For the $\Lambda$CCSD(T) triples correction $\delta E^{\rm (T)}$ itself, the $W_N$ contributions $\delta E_{\rm 3B}^{\rm (T)}$, despite containing second-order MBPT contributions, have very small values of about \mbox{--15\,keV}.
This effect is about one order of magnitude smaller than the targeted accuracy  given by the size of $\delta E^{\rm (T)}$ and may, therefore, be neglected. From another perspective, the $W_N$ contributions to $\delta E^{\rm (T)}$ constitute about \mbox{0.1\,\%} of the total binding energy, which clearly is beyond the level of accuracy of any many-body method operating in the medium-mass regime today.

As is apparent from Fig.~\ref{fig:IndividualContributions}, the situation for the $NN+3N$-induced Hamiltonian and the heavier nuclei \elem{O}{24} and \elem{Ca}{40} is similar. 
In the case of \elem{Ca}{40} we work in the smaller $\emax=10$ model space in order to keep the computational cost reasonable. In this model space the results are not fully converged with respect to \emax, but since the quality of the NO2B approximation is largely independent of $\emax$~\cite{BiLa13} this does not affect the present discussion. For the $NN+3N$-induced Hamiltonian, for example, the relative contribution of $W_N$ to the CCSD correlation energy grows from \mbox{1.3 \%} for $\alpha$ = 0.02\,\fm to \mbox{4.2\,\%} for $\alpha$ = 0.08\,\fmNoSpace, in both cases constituting  about \mbox{0.6\,\%} of the total binding energy. Again, as the SRG flow parameter increases, the contributions of $W_N$ to the CCSD correlation energy on the one hand, and the triples correction on the other hand, become comparable; $\Delta E^{\rm (CCSD)}- \Delta E_{\rm 2B}^{\rm (CCSD)}$ is about \mbox{18\,\%} of the size of the triples correction at $\alpha$ = 0.02\,\fm and already about \mbox{48\,\%} at $\alpha$ = 0.08\,\fmNoSpace.
The $W_N$ effect on the triples correction is again negligible, about one order of magnitude smaller than the triples correction itself, namely, about \mbox{2\,\%} of $\delta E^{\rm (T)}$ for $\alpha$ = 0.02\,\fm and about \mbox{11\,\%} for $\alpha$ = 0.08\,\fmNoSpace, or \mbox{0.1\,\%} and \mbox{0.2\,\%} of the total binding energy $E^{\rm ({\Lambda}CCSD(T))}$.

It should be noted that the apparent flow-parameter independence of 
$E^{\rm ({\Lambda}CCSD(T))}$ for the $NN+3N$-full Hamiltonian is accidental due to the $E_{3\max}$ cut used in our calculations. 
Increasing $E_{3\max}$ will move the energies upwards, and for the harder interactions it will do so to a larger extent than for the softer interactions. This leads to a reduction of the flow-parameter dependence of the $NN+3N$-induced results while the flow-parameter dependence of the $NN+3N$-full results is enhanced~\cite{BiLa13}..

In summary, for hard interactions, the residual 3$N$ effects to the CCSD correlation energy $E^{\rm (CCSD)}$ are rather small compared to the triples correction $\delta E^{\rm (T)}$, but they become comparable for soft interactions. Therefore, when using soft interactions, the residual 3$N$ interaction should be included in CCSD if the desired accuracy level also demands inclusion of triples excitation effects. For the triples correction, on the other hand, the residual 3$N$ interaction only plays an insignificant role, providing contributions that are shadowed by the considerably larger uncertainties stemming, e.g., from the cluster truncation. This motivates the use of the truncated energy expression 
$\tilde{E}^{\rm ({\Lambda}CCSD(T))}$, Eq.~(\ref{lambda-ccsd-t-partial}), instead of the full form $E^{\rm ({\Lambda}CCSD(T))}$, Eq.~(\ref{lambda-ccsd-t-full}),  resulting in only negligible losses in accuracy.

\begin{table*}												
\caption{Summary of the individual contributions to the $\Lambda$CCSD(T) ground-state energies in MeV for \elem{O}{16}, \elem{O}{24}, and \elem{Ca}{40} and for the $NN+3N$-induced and $NN+3N$-full Hamiltonians with $E_{3\max}=12$, obtained for \elem{O}{16} and \elem{O}{24} from an $\emax=12$ model space with oscillator frequency $\hbar \Omega=20$ MeV and for \elem{Ca}{40} from an $\emax=10$ model space with $\hbar \Omega=24$ MeV.
For the calculation of $\Delta E_{\rm 2B}^{\rm (CCSD)}$, amplitudes from the NO2B approximation have been used, while for the calculation of all other quantities the required amplitudes have been determined from equations including the residual 3$N$ interaction.
} 											
\label{tab:Data}												
\begin{ruledtabular}		
\begin{tabular}{cccccccc}												
  NN+3N-
& \multirow{2}{*}{$\alpha\,[\text{fm}^4]$}
& \multirow{2}{*}{$E^{\rm ({\Lambda}CCSD(T))}$}
& \multirow{2}{*}{$\ERef$}
& \multirow{2}{*}{$\Delta E_{\rm 2B}^{\rm (CCSD)}$}
& \multirow{2}{*}{$\Delta E^{\rm (CCSD)}-\Delta E_{\rm 2B}^{\rm (CCSD)}$} 
& \multirow{2}{*}{$\delta E_{\rm 2B}^{\rm (T)}$}
& \multirow{2}{*}{$\delta E_{\rm 3B}^{\rm (T)}$}
\\
induced & & & & & & & \\
\hline
\elem{O}{16} & 0.02 & --126.37 & --56.47 & --66.05 & 0.67 & --4.46 & --0.06 \\ 
 & 0.04 & --124.09 & --80.09 & --41.93 & 0.86 & --2.83 & --0.10 \\ 
 & 0.08 & --121.78 & --96.59 & --24.28 & 0.90 & --1.66 & --0.15 \\ 

\elem{O}{24} & 0.02 & --164.92 & --65.41 & --93.22 & 0.89 & --7.01 & --0.18 \\ 
 & 0.04 & --161.14 & --98.32 & --59.23 & 1.15 & --4.56 & --0.18 \\ 
 & 0.08 & --156.97 & --120.64 & --34.52 & 1.19 & --2.75 & --0.24 \\ 

\elem{Ca}{40} & 0.02 & --372.25 & --186.58 & --174.35 & 2.44 & --13.44 & --0.31 \\ 
 & 0.04 & --364.87 & --252.67 & --106.28 & 2.78 & --8.22 & --0.49 \\ 
 & 0.08 & --353.00 & --291.98 & --58.32 & 2.46 & --4.56 & --0.59 \\ 

\hline
  NN+3N-
& \multirow{2}{*}{$\alpha [\text{fm}^4]$}
& \multirow{2}{*}{$E^{\rm ({\Lambda}CCSD(T))}$}
& \multirow{2}{*}{$\ERef$}
& \multirow{2}{*}{$\Delta E_{\rm 2B}^{\rm (CCSD)}$}
& \multirow{2}{*}{$\Delta E^{\rm (CCSD)}-\Delta E_{\rm 2B}^{\rm (CCSD)}$}
& \multirow{2}{*}{$\delta E_{\rm 2B}^{\rm (T)}$}
& \multirow{2}{*}{$\delta E_{\rm 3B}^{\rm (T)}$}
\\
full & & & & & & & \\
\hline
\elem{O}{16} & 0.02 & --130.68 & --56.11 & --69.57 & 0.54 & --5.39 & --0.15 \\ 
 & 0.04 & --130.61 & --81.79 & --45.87 & 0.82 & --3.61 & --0.16 \\ 
 & 0.08 & --130.51 & --101.67 & --27.44 & 0.92 & --2.17 & --0.17
  \\ 

\elem{O}{24} & 0.02 & --171.28 & --64.16 & --99.53 & 0.67 & --8.01 & --0.25 \\ 
 & 0.04 & --171.82 & --101.52 & --65.81 & 1.07 & --5.28 & --0.28 \\ 
 & 0.08 & --171.65 & --130.43 & --39.01 & 1.18 & --3.05 & --0.35 \\ 

\elem{Ca}{40} & 0.02 & --369.56 & --158.28 & --194.80 & 2.12 & --17.80 & --0.80 \\ 
 & 0.04 & --375.02 & --238.62 & --126.64 & 2.96 & --11.86 & --0.86 \\ 
 & 0.08 & --375.82 & --298.75 & --72.23 & 2.85 & --6.82 & --0.87 \\ 
 
\end{tabular}
\end{ruledtabular}												
\end{table*}				

The above considerations indicate that the residual 3$N$ interaction may be neglected in calculating the $\Lambda$CCSD(T) energy correction $\delta E^{\rm (T)}$ without significantly affecting the overall accuracy, leading to Eq.~(\ref{lambda-ccsd-t-partial}) as an approximate form for $E^{\rm ({\Lambda}CCSD(T))}$.
From a practitioner's point of view,
discarding the $W_N$ contributions to $\delta E^{\rm (T)}$, Eqs.~(\ref{M-ijkabc-WN})--(\ref{ell-ijkabc-WN}), already leads to significant savings in the implementational effort and computing time which in our calculations requires about half a million CPU hours for one $\delta E^{\rm (T)}$ evaluation for \elem{O}{16} calculation at $\emax=12$ using full $W_N$ information with $E_{3\max}=12$, which is about two orders of magnitude more computationally expensive than the analogous calculation using the NO2B approximation.
However, one still has to solve the CCSD equations determining the $T^{\rm(CCSD)}$ amplitudes $t^a_i$ and $t^{ab}_{ij}$ as well as the $\Lambda$CCSD equations determining the $\Lambda^{\rm(CCSD)}$ amplitudes $\lambda^i_a$ and $\lambda^{ij}_{ab}$ with full incorporation of $W_N$. Particularly solving the $\Lambda$CCSD equations, for which the  similarity-transformed Hamiltonian contributions given in Tables~\ref{tab:EffectiveHamiltonian1B2B} and \ref{tab:EffectiveHamiltonian3B} have to be evaluated, consumes lots of the computing time in our calculations.
Therefore, it is also worthwhile to investigate how much of the residual 3$N$ information has to be incluced in solving for the amplitudes of the $T^{\rm(CCSD)}$ and $\Lambda^{\rm(CCSD)}$ operators that enter the energy expressions, in order to obtain accurate results at the lowest possible computational cost.

In order to distinguish different approximation schemes, we introduce the notation in which for energy quantities that only depend on $T^{\rm(CCSD)}$ amplitudes the label in brackets denotes if the $T^{\rm(CCSD)}$ amplitudes were determined from the amplitude equations with (3B) or without residual 3$N$ interaction (2B). Similarly, for  quantities that depend on both $T^{\rm(CCSD)}$ and $\Lambda^{\rm(CCSD)}$ amplitudes, the first label denotes the type of equation used to determine the $T^{\rm(CCSD)}$ amplitudes and the second one identifies the $\Lambda$CCSD equations. For example, $\tilde{E}^{\rm ({\Lambda}CCSD(T))}{\rm (3B,2B)}$ refers to the energy expression~(\ref{lambda-ccsd-t-partial}), calculated using $T^{\rm(CCSD)}$  amplitudes  determined from Eqs.~(\ref{momccsd1-NO2B}), (\ref{momccsd2-NO2B}),
(\ref{momccsd1-WN}) and (\ref{momccsd2-WN}) and the $\Lambda^{\rm(CCSD)}$ amplitudes  determined from Eqs.~(\ref{xi-ia-2B}) and (\ref{xi-ijab-2B}) only.

We consider the following approximation schemes, in which the $W_N$ contributions $\delta E_{\rm 3B}^{\rm (T)}$ to the triples correction are always neglected: For the ``NO2B'' scheme, all $W_N$ terms are discarded in both the determination of the  $T^{\rm(CCSD)}$ and $\Lambda^{\rm(CCSD)}$ amplitudes and the calculation of the energy $E_{\rm 2B}^{\rm ({\Lambda}CCSD(T))}$, Eq.~(\ref{lambda-ccsd-t-2B}),
\beq
E^{\rm(NO2B)} = E_{\rm 2B}^{\rm ({\Lambda}CCSD(T))} {\rm (2B,2B)} \ .
\label{appSchemeNO2B}
\eeq
This of course corresponds to an ordinary $\Lambda$CCSD(T) calculation in NO2B approximation.
For scheme ``A'', we compute $E_{\rm 2B}^{\rm ({\Lambda}CCSD(T))}$ as in the NO2B case and also add $\Delta E_{\rm 3B}^{\rm (CCSD)}$ with $T^{\rm(CCSD)}$  amplitudes obtained from the NO2B CCSD calculation,
\beq
E^{\rm (A)} = E_{\rm 2B}^{\rm ({\Lambda}CCSD(T))}{\rm (2B,2B)} + \Delta E_{\rm 3B}^{\rm (CCSD)} {\rm (2B)} \ .
\label{appSchemeA}
\eeq
This represents the simplest and most economic way to include $W_N$ information, where it only enters in the expression for the energy contribution $\Delta E_{\rm 3B}^{\rm (CCSD)}$, Eq.~(\ref{energy-WN}), but not in the considerably more complex equations that determine the amplitudes. 
In scheme ``B'', we include full $W_N$ information in the calculation of the CCSD correlation energy, keeping the $W_N$ terms in the amplitude equations as well as in the energy expression. The triples correction, however, is calculated without any $W_N$ information,
\beq
E^{\rm (B)} = E^{\rm (CCSD)}{\rm (3B)} + \delta E_{\rm 2B}^{\rm (T)}{\rm (2B,2B)}  \ .
\label{appSchemeB}
\eeq
This way we keep consistency between the $T^{\rm(CCSD)}$ and $\Lambda^{\rm(CCSD)}$ amplitudes that enter the triples correction, while capturing all residual 3$N$ effects in the CCSD energy $\Delta E^{\rm (CCSD)}$.
In scheme ``C'', we introduce an inconsistency between the $T^{\rm(CCSD)}$ and $\Lambda^{\rm(CCSD)}$ amplitudes by solving for $T^{\rm(CCSD)}$ with the $W_N$ terms present, 
while neglecting the $W_N$ terms in the equations for $\Lambda^{\rm(CCSD)}$, and calculate the energy using Eq.~(\ref{lambda-ccsd-t-partial}),
\beq
E^{\rm (C)} = \tilde{E}^{\rm ({\Lambda}CCSD(T))}{\rm (3B,2B)} \ .
\label{appSchemeC}
\eeq
This variant is reasonable since one typically has to solve for the $T^{\rm(CCSD)}$ amplitude equations with $W_N$ terms anyway in order to obtain the comparatively large $\Delta E_{\rm 3B}^{\rm (CCSD)}$ contribution to the energy while one would like to avoid to solve for the 
$\Lambda^{\rm(CCSD)}$ amplitudes in this manner, if the resulting energies do not change much.
Finally, in scheme ``D'', in which we only neglect the residual 3$N$ interaction terms in the expression for $\delta E^{\rm (T)}$, we use the full, $W_N$-containing equations to solve for the $T^{\rm(CCSD)}$ and $\Lambda^{\rm(CCSD)}$ amplitudes and determine the energy via Eq.~(\ref{lambda-ccsd-t-partial}),
\beq
E^{\rm (D)} = \tilde{E}^{\rm ({\Lambda}CCSD(T))}{\rm (3B,3B)} \ .
\label{appSchemeD}
\eeq
As in the discussion of Fig.~\ref{fig:IndividualContributions}, this variant allows to estimate the importance of $W_N$ for the $\Lambda^{\rm(CCSD)}$ amplitudes.

\begin{figure}[t]
\hspace{-20pt}
\includegraphics{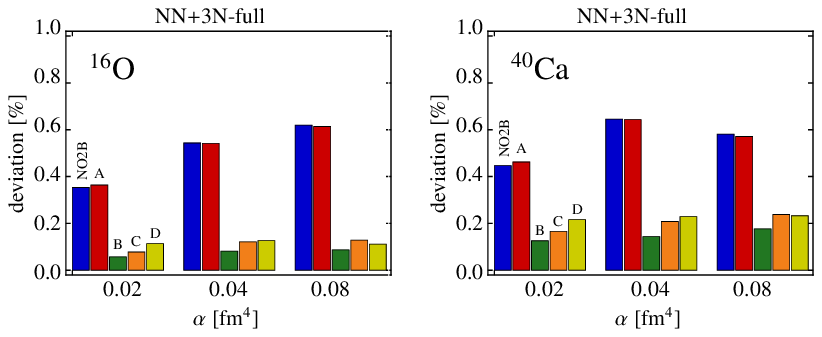} 
\vspace{-15pt}
\caption{
(color online) Comparison of the deviations introduced by the different approximation schemes, Eqs.~(\ref{appSchemeNO2B})--(\ref{appSchemeD}), described in the text from the full inclusion of the residual $3N$ interaction in all steps involving a CCSD and $\Lambda$CCSD(T) calculation.
}	
\label{fig:Approximations}
\end{figure}

In Fig.~\ref{fig:Approximations}, for the case of \elem{O}{16}, \elem{Ca}{40} and the $NN+3N$-full Hamiltonian, we compare the deviations of all the aforementioned approximation schemes from the complete 3$N$ calculations. For \elem{O}{24} and the $NN+3N$-induced Hamiltonian we obtain very similar results. As expected, the ``NO2B'' scheme shows the largest deviations because the contributions of $W_N$ to CCSD are completely missing. Including the $W_N$ terms in the energy expression for the CCSD correlation energy but evaluating it using $T^{\rm(CCSD)}$ amplitudes without $W_N$ information in scheme ``A'' virtually does not change the ``NO2B'' results. Therefore, we can conclude that it is the $W_N$ effect on the $T^{\rm(CCSD)}$ amplitudes that is most important for CCSD, and not the additional terms in $\Delta E_{\rm 3B}^{\rm (CCSD)}$.
In our calculations, the best approximation to the complete inclusion of residual 3$N$ interactions is provided by scheme ``B'', where we use full $W_N$ information to determine the CCSD correlation energy, but otherwise no $W_N$ information enters the calculation of the triples correction. However, approximation schemes ``B'',``C'', and ``D'' give very similar results, again hinting at the $W_N$ effect on the $T^{\rm(CCSD)}$ amplitudes to be the most important ingredient in the inclusion of residual 3$N$ interactions in CCSD and $\Lambda$CCSD(T) calculations.

%
%
\section{Conclusions}
\label{sec4}

In this article we considered the extension of CC theory with a full treatment of singly and doubly excited clusters and a non-iterative treatment of triply excited clusters to three-body Hamiltonians.
The incorporation of 3$N$ interactions into CCSD was previously discussed in detail in Ref.~\cite{HaPa07}, so in this article we focused on the corresponding generalization of the non-iterative treatment of triply excited clusters.
Among various triples corrections, for this first study we chose the $\Lambda$CCSD(T)-type approach due to its relatively simple structure.

The $\Lambda$CCSD(T) approach requires one to solve the $\Lambda$CCSD equations prior to the computation of the actual energy correction. Thus, in addition to the explicit energy expressions defining the $\Lambda$CCSD(T) approach for three-body Hamiltonians, we also provided a detailed discussion of the inclusion of 3$N$ interactions into the $\Lambda$CCSD equations, listing the complete set of the relevant programmable expressions. 
The similarity-transformed Hamiltonian is a central quantity of coupled-cluster theory and in this article we give explicit expressions for the contributions of the residual 3$N$ interactions to all one- and two-body components as well as selected three- and four-body components of this Hamiltonian.
We derived the $\Lambda$CCSD(T) method as an approximation to the more complete CR-CC(2,3) approach which allows for an easy identification of new terms arising due to the presence of residual 3$N$ interactions, and we provided complete and explicit expressions required in the calculation of the $\Lambda$CCSD(T) energy correction for three-body Hamiltonians. 

One of the important outcomes of our analysis is the realization that through the use of explicit 3$N$ interactions in $\Lambda$CCSD(T), compared to the approximate NO2B treatment, contributions of the triply excited clusters are moved from second to first order in MBPT for the wave function, and from fourth to second order for the energy.
This is rather easy to account for at the full CCSDT level, which is, unfortunately prohibitively expensive, but is not trivial at all when one tries to account for the $T_3$ cluster contributions via corrections to the CCSD energy.
The use of the CR-CC(2,3) formalism, which contains the $\Lambda$CCSD(T) approach as an approximation, turned out to be central for properly accounting for the second-order MBPT corrections due to the $T_3$ clusters induced by residual 3$N$ interactions and other related terms.

The method was applied to the medium-mass closed-shell nuclei \elem{O}{16}, \elem{O}{24} and \elem{Ca}{40} using $NN+3N$ Hamiltonians obtained from chiral EFT. For the total binding energies, the effect of the residual three-body interactions at the level of CCSD can become comparable to the $\Lambda$CCSD(T) correction, particularly for soft interactions, while
for the $\Lambda$CCSD(T) correction itself, contributions of the residual 3$N$ interactions were shown to be negligible.
Therefore, for the CCSD and $\Lambda$CCSD(T) calculations, by only including explicit 3$N$ interactions at the CCSD level, we can practically eliminate the error introduced by the normal-ordering approximation.
We further discussed various combinations of where to include the residual 3$N$ interactions in the determination of the amplitudes from which energies are calculated, and found that the residual 3$N$ interactions have their most significant effect on the cluster amplitudes, i.e., it is important to solve the CCSD equations including residual 3$N$ interactions when determining the CCSD energy, but one can safely neglect these interactions in post-CCSD corrections due to $T_3$ clusters.

\section{Acknowledgments}

Numerical calculations have been performed at the J\"ulich Supercomputing Centre, at the LOEWE-CSC Frankfurt, and at the National Energy Research Scientific Computing Center supported by the Office of Science of the U.S. Department of Energy under Contract No.~DE-AC02-05CH11231.  
Supported by the Deutsche Forschungsgemeinschaft through contract SFB 634, by the Helmholtz International Center for FAIR within the LOEWE program of the State of Hesse, and the BMBF through contracts 06DA9040I and 06DA7047I. Additional support by the Chemical Sciences, Geosciences and Biosciences Division, Office of Basic Energy Sciences, Office of Science, U.S. Department of Energy (Grant No. DE-FG02-01ER15228 to P.P.) is gratefully acknowledged. P.N. acknowledges support from Natural Sciences and Engineering Research Council of Canada (NSERC) Grant No. 401945-2011.


		
\end{document}